\documentclass{pasj01}

\begin{document}

\author{Mariko \textsc{Kimura}\altaffilmark{1,*}, 
        Shinya \textsc{Yamada}\altaffilmark{2}, 
        Nozomi \textsc{Nakaniwa}\altaffilmark{3}, 
        Yoshihiro \textsc{Makita}\altaffilmark{2}, 
        Hitoshi \textsc{Negoro}\altaffilmark{4}, 
        Megumi \textsc{Shidatsu}\altaffilmark{5}, 
        Taichi \textsc{Kato}\altaffilmark{6}, 
        Teruaki \textsc{Enoto}\altaffilmark{1}, 
        Keisuke \textsc{Isogai}\altaffilmark{7,8}, 
        Tatehiro \textsc{Mihara}\altaffilmark{1}, 
        Hidehiko \textsc{Akazawa}\altaffilmark{9}, 
        Keith C.~\textsc{Gendreau}\altaffilmark{10}, 
        Franz-Josef~\textsc{Hambsch}\altaffilmark{11,12,13}, 
        Pavol A.~\textsc{Dubovsky}\altaffilmark{14}, 
        Igor~\textsc{Kudzej}\altaffilmark{14}, 
        Kiyoshi \textsc{Kasai}\altaffilmark{15}, 
        Tam\'{a}s \textsc{Tordai}\altaffilmark{16}, 
        Elena \textsc{Pavlenko}\altaffilmark{17,18}, 
        Aleksei A.~\textsc{Sosnovskij}\altaffilmark{17}, 
        Julia V.~\textsc{Babina}\altaffilmark{17}, 
        Oksana I.~\textsc{Antonyuk}\altaffilmark{17}, 
        Hiroshi~\textsc{Itoh}\altaffilmark{19}, and
        Hiroyuki~\textsc{Maehara}\altaffilmark{7,20,21}
        }
\email{mariko.kimura@riken.jp}

\altaffiltext{1}{Cluster for Pioneering Research, Institute of Physical and Chemical Research (RIKEN), 2-1 Hirosawa, Wako, Saitama 351-0198}
\altaffiltext{2}{Department of Physics, Rikkyo University, 3-34-1 Nishi-Ikebukuro, Toshima-ku, Tokyo 171-8501}
\altaffiltext{3}{Department of Physics, Tokyo Metropolitan University, 1-1 Minami-Osawa, Hachioji, Tokyo 192-0397}
\altaffiltext{4}{Department of Physics, Nihon University, 1-8 Kanda-Surugadai, Chiyoda-ku, Tokyo 101-8308}
\altaffiltext{5}{Department of Physics, Ehime University, 2-5, Bunkyocho, Matsuyama, Ehime 790-8577}
\altaffiltext{6}{Department of Astronomy, Graduate School of Science, Kyoto University, Oiwakecho, Kitashirakawa, Sakyo-ku, Kyoto 606-8502}
\altaffiltext{7}{Okayama Observatory, Kyoto University, 3037-5 Honjo, Kamogatacho, Asakuchi, Okayama 719-0232}
\altaffiltext{8}{Department of Multi-Disciplinary Sciences, Graduate School of Arts and Sciences, The University of Tokyo, 3-8-1 Komaba, Meguro, Tokyo 153-8902}
\altaffiltext{9}{Akazawa Funao Observartory, 107 Funao, Funaocho, Kurashiki, Okayama 710-0261}
\altaffiltext{10}{Astrophysics Science Division, NASA Goddard Space Flight Center, Greenbelt, MD 20771, USA}
\altaffiltext{11}{Groupe Europe\'{e}n d'Observations Stellaires (GEOS), 23 Parc de Levesville, 28300 Bailleau l'Ev\^{e}que, France}
\altaffiltext{12}{Bundesdeutsche Arbeitsgemeinschaft f\"{u}r Ver\"{a}nderliche Sterne (BAV), Munsterdamm 90, 12169 Berlin, Germany}
\altaffiltext{13}{Vereniging Voor Sterrenkunde (VVS), Oude Bleken 12, 2400 Mol, Belgium}
\altaffiltext{14}{Vihorlat Observatory, Mierova 4, 06601 Humenne, Slovakia}
\altaffiltext{15}{Baselstrasse 133D, CH-4132 Muttenz, Switzerland}
\altaffiltext{16}{Polaris Observatory, Hungarian Astronomical Association, Laborc utca 2/c, 1037 Budapest, Hungary}
\altaffiltext{17}{Federal State Budget Scientific Institution ``Crimean Astrophysical Observatory of RAS'', Nauchny, 298409, Republic of Crimea}
\altaffiltext{18}{V.~I.~Vernadsky Crimean Federal University, 4 Vernadskogo Prospekt, Simferopol, 295007, Republic of Crimea}
\altaffiltext{19}{Variable Star Observers League in Japan (VSOLJ), 1001-105 Nishiterakata-machi, Hachioji, Tokyo 192-0153}
\altaffiltext{20}{Subaru Telescope Okayama Branch Office, National Astronomical Observatory of Japan, National Institutes of Natural Sciences, 3037-5 Honjo, Kamogata, Asakuchi, Okayama 719-0232}
\altaffiltext{21}{Variable Star Observers League in Japan (VSOLJ), Okayama, Japan}

\title{
On the nature of the anomalous event in 2021 in the dwarf nova SS Cygni and its multi-wavelength transition
}

\Received{} \Accepted{}


\SetRunningHead{Kimura et al.}{First standstill-like phenomenon in SS Cyg}

\maketitle

\begin{abstract}

SS Cyg has long been recognized as the prototype 
of a group of dwarf novae that show only outbursts.  
However, this object has entered a quite anomalous event 
in 2021, which at first appeared to be standstill, i.e., 
an almost constant luminosity state, observed 
in Z Cam-type dwarf novae.  
This unexpected event gives us a great opportunity to 
reconsider the nature of standstill in cataclysmic variables.  
We have observed this anomalous event and its forerunner, 
a gradual and simultaneous increase in the optical 
and X-ray flux during quiescence, through 
many optical telescopes and the X-ray telescopes {\it NICER} 
and {\it NuSTAR}.  
We have not found any amplification of the orbital hump 
during quiescence before the anomalous event, \textcolor{black}{which 
suggests that the mass transfer rate did not significantly 
fluctuate on average}.  
The estimated X-ray flux was not enough to explain 
the increment of the optical flux during quiescence 
via X-ray irradiation of the disk and the secondary star.  
It would be natural to consider that viscosity in 
the quiescent disk was enhanced before the anomalous event, 
which increased mass accretion rates in the disk and raised 
not only the optical flux but also the X-ray flux.  
We suggest that enhanced viscosity also triggered 
the standstill-like phenomenon in SS Cyg, which is considered 
to be a series of small outbursts.  
The inner part of the disk would always stay in the outburst 
state and only its outer part would be unstable against 
the thermal-viscous instability during this phenomenon, 
which is consistent with the observed optical color variations.  
This scenario is in line with our X-ray spectral analyses 
which imply that the X-ray emitting inner accretion flow became 
hotter than usual and vertically expanded and that it became denser 
and was cooled down after the onset of the standstill-like state.  

\end{abstract}

\section{Introduction}

Cataclysmic variables (CVs) are close binary systems 
composed of a primary white dwarf (WD) and 
a secondary low-mass cool star.
The secondary star provides its mass to the disk 
by Roche-lobe overflow and an accretion disk is 
formed around the WD.
Dwarf novae (DNe), one subclass of CVs, show 
intermittent outbursts which are the sudden 
brightening of the disk 
(for a general review, see \cite{war95book}).  

It is widely accepted that thermal-viscous instability 
triggered by partial ionization of hydrogen causes 
dwarf-nova outbursts, and this model is called 
``the disk-instability model'' 
(\cite{can93review,osa96review,ham20review} for reviews).  
In this model, the thermal equilibrium curve of 
a given radius of the disk has an unstable state 
sandwiched by two stable states: the hot state 
corresponding to the outburst state with a high 
accretion rate and the cool state corresponding to 
the quiescent state with a low accretion rate 
\citep{mey81DNoutburst}. 
If a local region of the disk jumps between these two 
stable states, transition waves propagate and alter 
the thermal state of the entire disk at once 
(e.g., \cite{sma84DI,min85DNDI}).

The basic assumption in the disk-instability model is 
that the mass transfer rate from the secondary star 
is constant.  
The mass transfer rate is one of 
the system parameters for the classification of DNe.  
If it is less than the critical rate denoted as 
$\dot{M}_{\rm crit}$ above which the disk is thermally stable, 
the systems repeat outbursts and are called SS Cyg-type stars.
If the mass transfer rate exceeds $\dot{M}_{\rm crit}$, 
the systems show no eruptive events and are called 
nova-like stars (NLs).  
If the mass transfer rate 
\textcolor{black}{is close to and fluctuates around} 
$\dot{M}_{\rm crit}$, 
the systems alternate between frequent dwarf-nova outbursts 
and standstill, i.e., a state having constant luminosity 
in between that of outburst and quiescence.  
They are named Z Cam-type stars and regarded as 
the intermediate class between SS Cyg-type stars and NLs.  

The alternation between standstill and outbursts 
in Z Cam-type stars cannot be understood by the simple 
disk-instability model 
and it is believed that fluctuations of transfer 
rates are necessary (e.g., \cite{mey83zcam,bua01zcam}).  
\citet{ros17zcam} attempted to reproduce the Z Cam-type 
phenomenon by fluctuations of the turbulent viscosity 
without variations in mass transfer rates, but they failed 
to generate long-lasting standstills and several 
consecutive outbursts.  
We thus have not yet explained all of the dwarf-nova 
outbursts by the simple disk-instability model, so that 
the disk-instability model should be sublimated towards 
a truly unified model that can explain a wide variety 
of light variations in DNe.  

SS Cyg is the brightest DN over various wavelengths.
This system has been monitored by visual observations and 
optical telescopes for more than 100 years (see the AAVSO 
historical light curve\footnote{$<$https://www.aavso.org/sites/default/files/images/historicLC-SSCyg.jpg$>$}) and has constantly repeated 
normal dwarf-nova outbursts with amplitudes of $\sim$3.5 mag 
and with intervals of $\sim$1 month in the long history of 
optical observations.  
This source thus has been the prototype of SS Cyg-type stars.  
However, it has entered a state of anomaly that 
we have never seen before since the end of January 2021.  
This event seemed to be a genuine standstill at first sight 
(vsnet-alert 25453) and was completely unexpected.  
SS Cyg may be no longer the prototype of SS Cyg-type stars.  

This anomalous event in SS Cyg was accompanied with 
its predecessor phenomenon, which is the gradual and simultaneous 
increase in the optical and X-ray flux in the quiescent state.  
This phenomenon started around August 2019.
The X-ray monitoring has been continued for more than 
20 years\footnote{$<$http://maxi.riken.jp/pubdata/v3/J2142+435/index.html$>$ and $<$http://xte.mit.edu/asmlc/srcs/sscyg.html$>$}, 
but the long-lasting increase of the system brightness 
was the first event at X-rays.  
We noticed this multi-wavelength transition towards 
the anomalous event in 2021 and have monitored this source 
at optical and X-ray wavelengths 
since 2020 via the Variable Star Network (VSNET) collaboration 
team, the American Association of Variable Star Observers 
(AAVSO) \citep{waa20sscyg}, the Neutron star Interior Composition 
ExploreR ({\it NICER}; \cite{NICER}), and the Nuclear Spectroscopic 
Telescope Array ({\it NuSTAR}; \cite{NuSTAR}).  

It is not obvious that SS Cyg in which the mass transfer rate 
is not close to the critical rate exhibits a standstill-like 
phenomenon.  
The present event in SS Cyg may therefore give some 
important suggestions to the long-standing question 
in the study of dwarf-nova outbursts 
``are mass-transfer rate 
fluctuations necessary to reproduce standstill ?'' 
and we may have to reconsider the nature of standstill in CVs.  
Our primary goal is to understand what causes the unexpected 
event and its transition in SS Cyg by analyzing multi-wavelength data.
Simultaneously, the increase in the X-ray flux gives us a good 
opportunity to explore the inner accretion flow which is 
not well investigated (see \cite{bal20xrayCV} for a review).  
Section~2 describes our optical and X-ray observations.  
Section~3 shows the results of our data analyses.  
We discuss the cause of the anomalous event in 2021 and 
its multi-wavelength transition and provide the entire accretion 
picture in section~4.  A summary is given in section~5.

\section{Observations and data reduction}

\subsection{Optical photometry}

Time-resolved CCD photometric observations of SS Cyg were 
carried out by the VSNET collaboration team and the AAVSO.  
The log of the observation by the VSNET team is given in Table E1.  
We use the AAVSO archive data\footnote{$<$http://www.aavso.org/data/download/$>$} 
in the $B$, $V$, $R_{\rm C}$ and $I_{\rm C}$ bands after 
BJD 2453396.  
We convert all of the observation times to barycentric 
Julian date (BJD) in terrestrial time (TT).  
The data reduction and the calibration of the comparison stars 
were performed by each observer.  
The constancy of each comparison star was checked by nearby 
stars in the same images.  
The magnitude of each comparison star was measured 
by the AAVSO Photometric All-Sky Survey 
(APASS: \cite{APASS}) from the AAVSO Variable Star 
Database\footnote{$<$http://www.aavso.org/vsp$>$}.

\subsection{NICER}

{\it NICER} has monitored SS Cyg since May 2020.  
We use part of the data taken in optical quiescence 
during the 2021 anomalous event and its pre-stage, whose 
periods are defined in subsection 3.1.  The corresponding 
ObsIDs are given in Table E2.  
In this work, we utilize HEAsoft version 6.27.2 for data 
reduction and analyses.  
The data were reprocessed with the pipeline tool \texttt{nicerl2} 
based on the NICER Calibration Database (CALDB) version 
later than 2019 May 16, 
before producing light curves and time-averaged spectra 
of the individual ObsIDs.  
The background spectra are extracted by \texttt{nibackgen3C50} 
version 6.  
For spectral analyses, we obtained the response matrix file 
and the ancillary response file for a specific set of 
50 detectors to match the default settings of the background 
model\footnote{The method is described in $<$https://heasarc.gsfc.nasa.gov/docs/nicer/analysis\_threads/arf-rmf/$>$ and we use the additional data 
version xti20200722.}.

\subsection{NuSTAR}

The {\it NuSTAR} Target of Opportunity (ToO) observations 
were coordinated by some of the {\it NICER} observations.  
The ObsIDs of these observations that we use in this paper 
are given in Table E3.  
The data are reprocessed through \texttt{nupipeline} and 
the {\it NuSTAR} CALDB as of \textcolor{black}{2021 April 27}.  
The light curves, time-averaged spectra, and response and 
ancillary response files are obtained by \texttt{nuproducts}.  
\textcolor{black}{If the source is bright, we sometimes need 
the Multi Layer Insulation (MLI) correction \citep{mad20MLI} 
for the FPMA data\footnote{$<$https://nustarsoc.caltech.edu/NuSTAR\_Public/NuSTAROperationSite/mli.php$>$}.  
We extracted the FPMA spectra of the ObsIDs 90701304002, 
90702309002, and 90702309004 by using the old ancillary 
response file.}  
The background region is extracted as a circular region with 
a radius of 80'' at a blank sky area.  
We determine the source region as a circular 
region centered on the target with a 100--120'' radius 
by considering the time-varying brightness of the target.

\ifnum0=1
\subsection{MAXI}

\textcolor{blue}{
MAXI/GSC monitors SS Cyg.  
We retrieved the GSC light curves through the on-demand 
system\footnote{http://maxi.riken.jp/mxondem/}.}  
\fi

\section{Results of optical and X-ray analyses}

\subsection{Overall optical and X-ray light curves}

\begin{figure*}[htb]
\vspace{-5mm}
\begin{center}
\FigureFile(160mm, 50mm){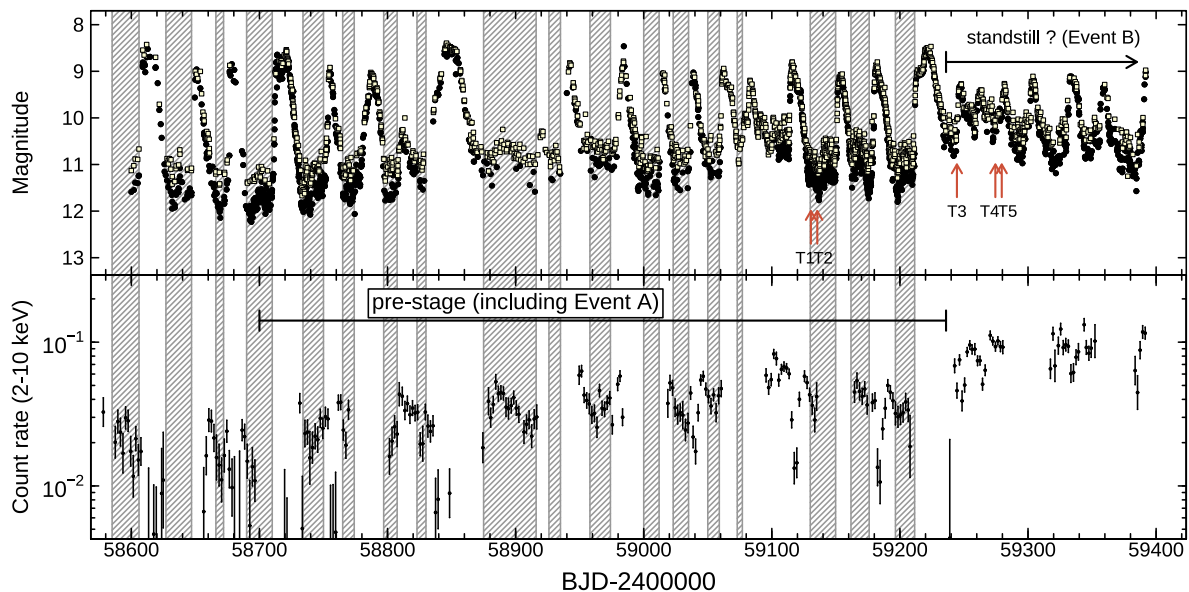}
\end{center}
\caption{
Long-term optical and X-ray light curves.  The upper panel exhibits optical light curves taken by the AAVSO and the VSNET observers.  The photometric data are binned to 0.1~d.  The circles and squares stand for the $V$-band and $R_{\rm C}$-band photometry, respectively.  The lower panel shows the {\it MAXI} light curve in the 2--10 keV, respectively.  The data are binned to 2~d.  We denote the time periods of the optical quiescence by the shaded regions.  In the upper panel, the arrows indicate the dates when coordinated observations with {\it NICER} and {\it NuSTAR} were performed, respectively.  
}
\label{multi-overall}
\end{figure*}

First of all, we show the long-term optical and X-ray 
light curves in Figure \ref{multi-overall}.  
The X-ray light curves that we here use are 
Monitor of All-sky X-ray Image (MAXI; \cite{mat09MAXI}) / 
Gas Slit Camera (GSC) light curves, which are retrieved 
through the on-demand 
system\footnote{$<$http://maxi.riken.jp/mxondem/$>$}.  
In this figure, we indicate the time periods of 
the optical quiescent state by the shaded regions.  
We see that the optical and X-ray flux began increasing 
around BJD 2458700 in the optical quiescent state.  
\textcolor{black}{The $R_{\rm C}$-band light curve only during 
the quiescent state is given in Fig.~E1}.  
The mean flux in the quiescent state seems to have 
continuously increased at two wavelengths until the onset 
of the long outburst triggered on BJD 2459210, 
though it modulated on timescales of $\sim$100~d.  
After the long outburst, the system showed small amplitude 
fluctuations.  
We define the time period between BJD 2458700 and 
BJD 2459236 as the pre-stage and \textcolor{black}{the gradual increase 
in the optical and X-ray fluxes in the hatched time intervals 
of the optical quiescent state during the pre-stage as Event A}.  
We also define the anomalous event which is possibly 
similar to standstill after BJD 2459236 as Event B.  

The optical luminosity increased by $\sim$1 mag during 
Event A.  
The optical outbursts observed in SS Cyg are roughly 
classified into two types: short outbursts with small 
amplitudes and long outbursts with large amplitudes.  
These outbursts show steep rises and \textcolor{black}{they are considered to 
be triggered at the outer disk}, i.e., outside-in outbursts, 
and the long outburst and the short outburst are alternately 
repeated \citep{can98sscyg}.  
The slow-rise outbursts are \textcolor{black}{likely} inside-out outbursts, 
which are triggered at the inner region of the disk, and 
relatively rare events (see the AAVSO historical light curve).  
However, the slow-rise and small-amplitude outbursts 
frequently occurred especially after BJD 2458700.  
Also, anomalous outbursts with very low amplitudes sometimes 
occurred in the pre-stage.  
The X-ray flux temporarily drops in the optical outburst 
state, which is consistent with the observations reported 
by \citet{whe03sscyg} and \citet{mcg04sscygXray}.  

\begin{figure}[htb]
\begin{center}
\FigureFile(80mm, 50mm){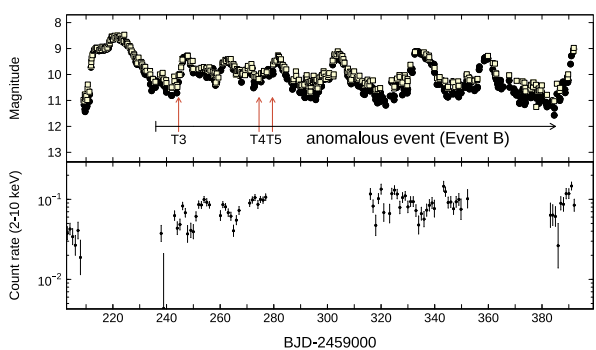}
\end{center}
\caption{
Recent activity of SS Cyg.  
These panels are the same as the upper and lower panels of Figure \ref{multi-overall} but after the onset of the 2021 long outburst.  
In the lower panel, the data are binned to 1~d.  
}
\label{multi-standstill}
\end{figure}

We next focus on the light curves around Event B 
in Figure \ref{multi-standstill}.  
SS Cyg entered a precursor outburst on BJD 2459210.  
Its precursor has a shoulder-like shape and lasted for 
$\sim$5~d before the outburst maximum.  
Although the precursor is sometimes observed at the beginning 
of long outbursts in SS Cyg-type stars 
\citep{can12ugemLC,ram12v447lyr}, the duration is mostly 
a couple of days.  
The precursor in this outburst is therefore extremely long.  
Also, we do not see a clear plateau stage in this long outburst, 
though the long outburst in SS Cyg normally has 
a gradually-decaying plateau stage during $\sim$1 week before 
the fading to the quiescent level.  
This long outburst faded around BJD 2459230 and 
SS Cyg entered a standstill-like phenomenon with oscillatory 
variations since BJD 2459236.  
Although a similar event might occur in 1908 in this source, 
the period when the quiescent level was rising is around 100~d, 
which is shorter than that of the pre-stage, and the anomalous 
event in 1908 looks like a group of $\sim$2-mag-amplitude 
small outbursts (see Figure E2).  
\textcolor{black}{Also, a similar multi-wavelength behavior was 
seen around BJD 2459100.  
However, the event after BJD 2459236 is more anomalous and 
long-lasting.}  
The optical luminosity is considered to approximately 
represent the gravitational energy released from the disk 
via mass accretion.  
We average the $V$-band light curves per day and 
derive the mean luminosity before Event B to be 
$\sim$10.3 mag.  
The mean luminosity in Event B is $\sim$10.1~mag, 
which is comparable with that before Event B\footnote{\textcolor{black}{Here, the mean luminosity is averaged over fluxes and converted into magnitudes.}}.  

\subsection{Optical orbital light curves}

\begin{figure*}[htb]
\begin{center}
\begin{minipage}{0.49\hsize}
\FigureFile(80mm, 50mm){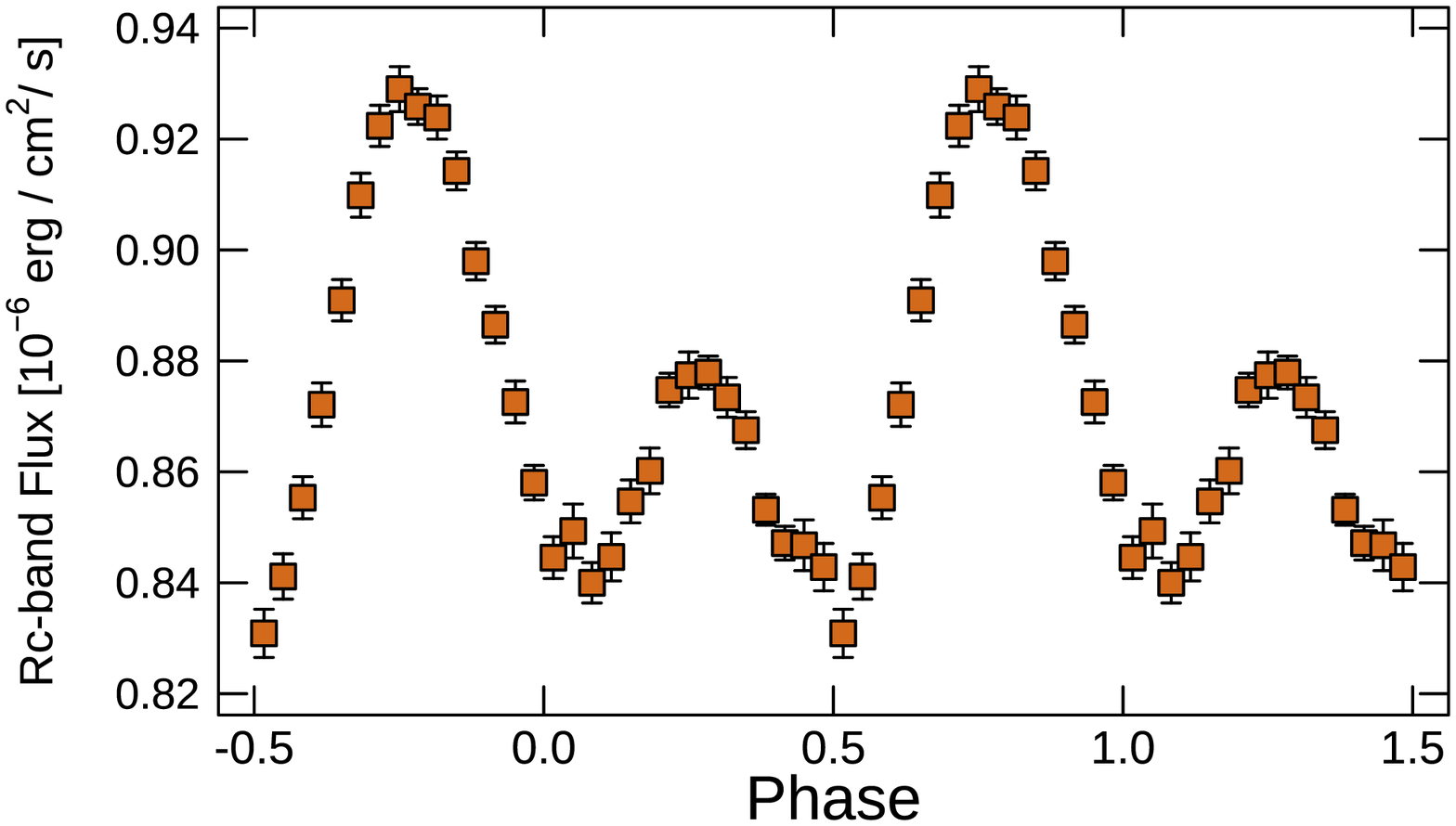}
\end{minipage}
\begin{minipage}{0.49\hsize}
\FigureFile(80mm, 50mm){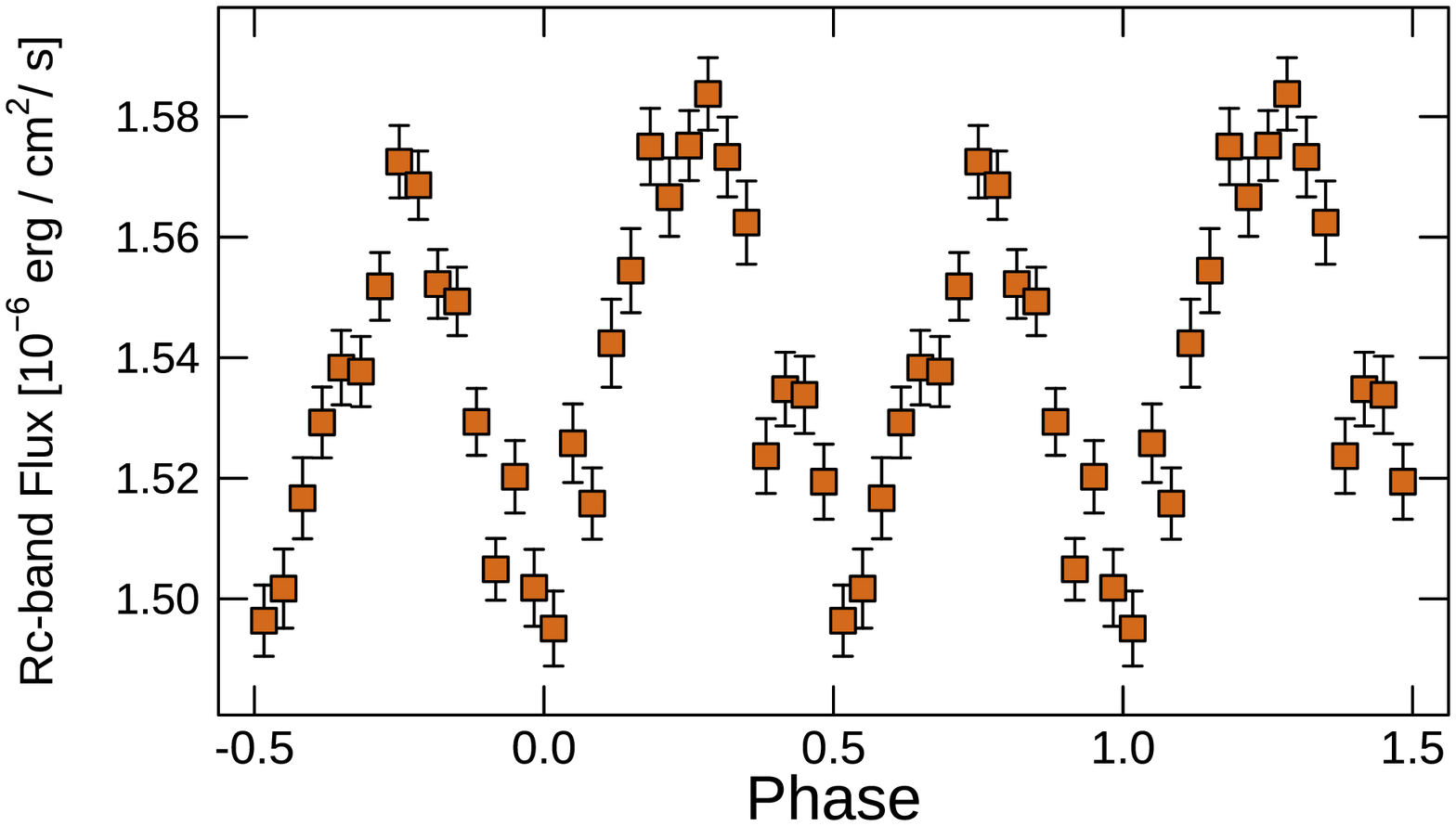}
\end{minipage}
\end{center}
\caption{Orbital light curves in the $R_{\rm C}$ band.  The left and the right panels are the orbital profile in the optical quiescence before the pre-stage and that in Event A, respectively.  }
\label{rband-orbital}
\end{figure*}

We here show the averaged $R_{\rm C}$-band orbital light 
curve in the optical quiescence before the pre-stage 
and that in Event A in Figure \ref{rband-orbital}.  
For the data before the pre-stage, we extract 
the light curve fainter than 11 mag 
\textcolor{black}{during BJD 2453868--2458700}.  
We fold the $R_{\rm C}$-band light curve with the orbital period 
of 0.27512973~d derived by \citet{hes86sscyg} and determine 
the orbital phase using the epoch BJD 2456190.62771 
reported by \citet{hil17sscyg}, which corresponds to 
the inferior conjunction of the secondary star.  
Before folding light curves, we subtract the long-term trend 
of light curves from the observational data by locally 
weighted polynomial regression (LOWESS; \cite{LOWESS}).  

The orbital light curve in the optical quiescence before 
the pre-stage, which is exhibited in the left panel 
of Figure \ref{rband-orbital}, can be interpreted 
as the superposition of the orbital hump and 
the ellipsoidal modulation.  
The orbital hump originates from the bright spot which is 
created at the disk outer rim by the collision between 
the gas stream from the secondary star and the disk and 
appears around the orbital phase of $\sim$0.8.  
This is because the bright spot rotates with the binary 
orbital motion and is hidden from the observer 
when the secondary star is behind the optically-thick disk 
(see e.g., \cite{woo86zcha}).  
The ellipsoidal modulation is generated by the change in 
the projection area of the distorted secondary star to 
the observer and peaks at the orbital phases of 0.25 and 0.75 
(see e.g., \cite{all96xyari}).  
Since the inclination angle of SS Cyg is measured to be 45 deg 
\citep{hil17sscyg}, these two effects are observable and 
the disk and the WD are not eclipsed.  

Comparing the left and the right panels of 
Figure \ref{rband-orbital}, we see that \textcolor{black}{the flux amplitude 
of the hump with a peak around the phase 0.8, which is 
measured from the light minimum around the phase 0.0, does not 
significantly change.  Although both the ellipsoidal modulation 
and the orbital hump contribute to this hump, we find 
no observational evidence that the luminosity of 
the secondary star largely changes during Event A (see also 
subsection~4.1).  
Therefore, the flux amplitude in the orbital hump would 
not change on average during Event A, 
which means that the bright spot in Event A is 
as bright as that in the quiescent state before 
the pre-stage.}  
\textcolor{black}{On the other hand, the hump around the orbital phase 0.3 
in the right panel may originate from X-ray irradiation of 
the secondary star.  
As estimated in subsection~4.1, the temperature of 
the irradiated surface of the secondary star increases by 
a few percent at most in Event A.  
If we take into account that part of the irradiated surface is 
not seen to the observer around the phase 0.0, 
$\sim$0.1-mag hump is possible at most.  This is enough to 
explain the amplitude of this hump.  
However, it is not straightforward to understand that 
the peak phase of this hump deviates from 0.5.  
Although this phenomenon was also detected in another CV \citep{kim20kic940}, 
its interpretation is unclear.  
One possibility is that the part of the outer edge of 
the disk that collides with the gas stream from 
the secondary star, i.e., the location of the bright spot, 
is geometrically thicker than the rest of the outer edge 
of the disk, making the irradiated surface of the secondary 
star asymmetric with respect to the line connecting the WD 
and the secondary star.  
In this case, part of the X-ray radiation would be interrupted 
by the thicker part and the observable area of the irradiated 
surface of the secondary star would be maximized at the orbital 
phase earlier than 0.5.}

\subsection{Optical color variations}

\begin{figure*}[htb]
\begin{center}
\begin{minipage}{0.49\hsize}
\FigureFile(80mm, 50mm){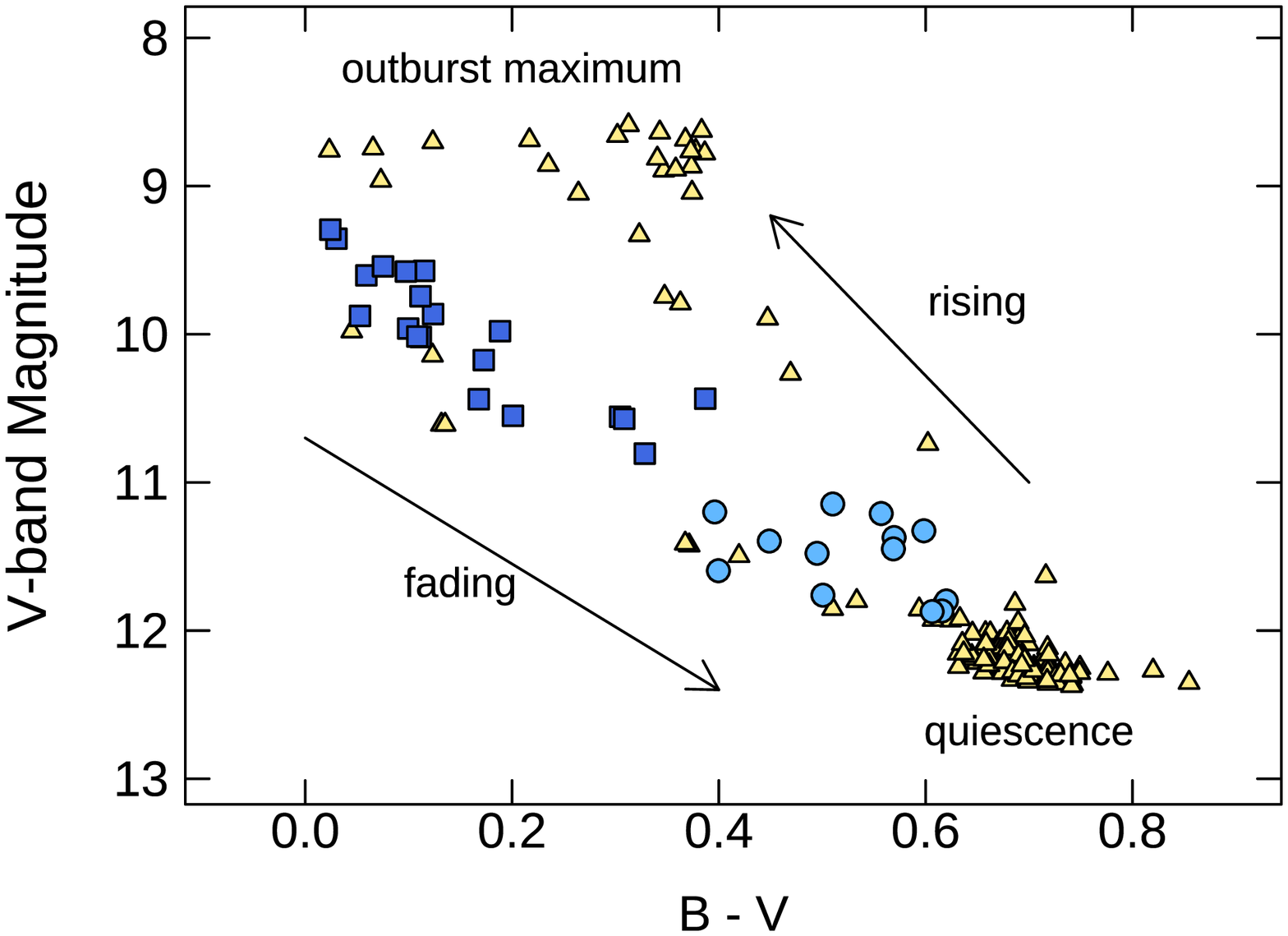}
\end{minipage}
\begin{minipage}{0.49\hsize}
\FigureFile(80mm, 50mm){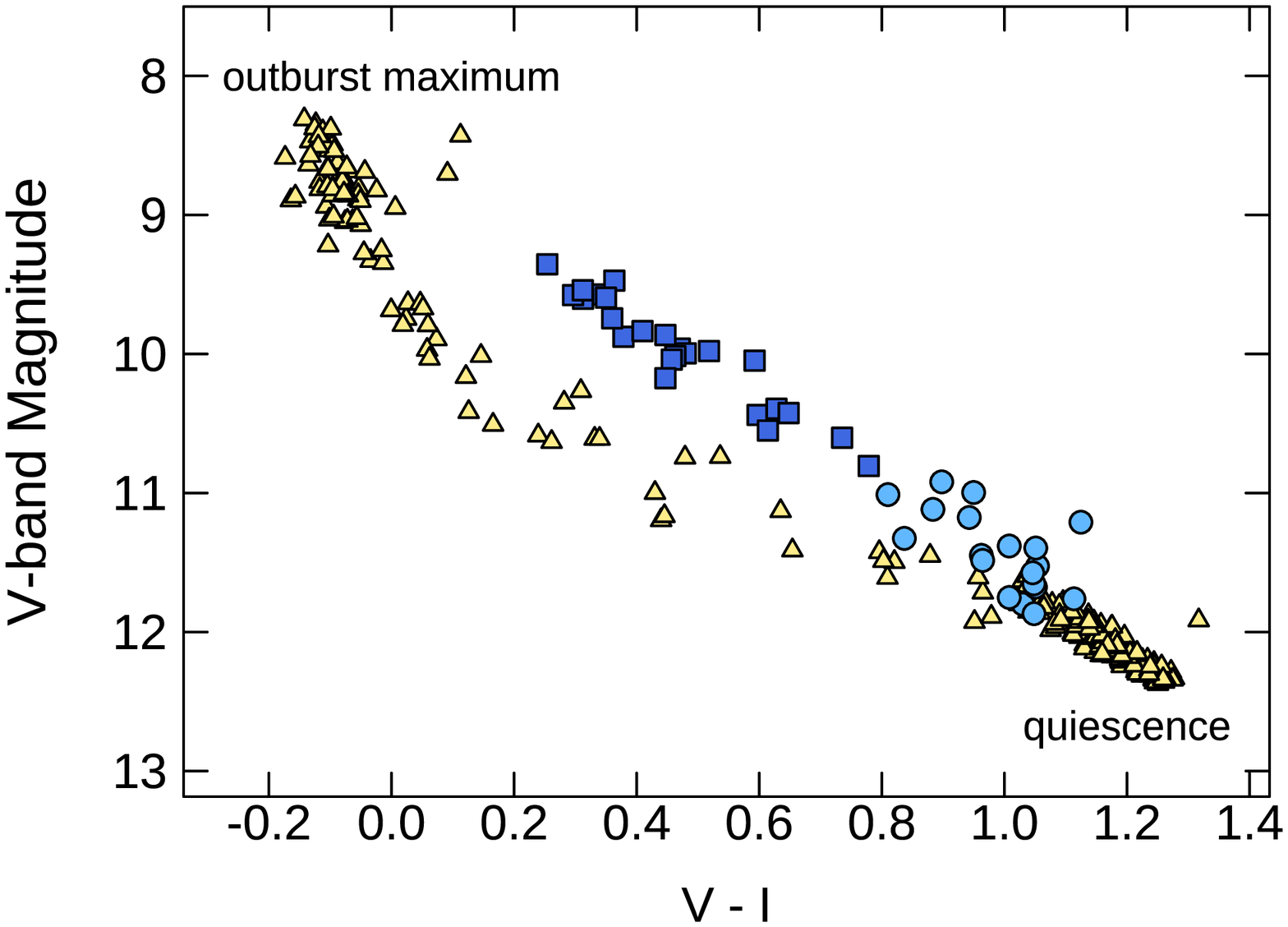}
\end{minipage}
\end{center}
\caption{
Optical color variations with respect to the $V$-band magnitude.  The left and right panels show the $B - V$ and $V - I$ color variations, respectively.  The triangles, circles, and squares represent the data during normal long outbursts between BJD 2456130 and BJD 2456280, those in Event A, and those in Event B, respectively.  
}
\label{color}
\end{figure*}

We here explore the optical color variations.  
We average the light curves per orbital cycle and 
extract the colors.  
We select the data of the two normal long outbursts 
between BJD 2456130 and BJD 2456280 for this analysis 
as did in Fig.~2 of \citet{ham20modeling} 
and compare the color variations in Event A and Event B.  
The 1$\sigma$ errors are typically around 0.1~mag 
and we exclude the data with errors greater than 0.2~mag.  
\textcolor{black}{
Also, we use the simultaneous observation data and 
exclude the data with the mean orbital phase of 
0.75--0.85 to avoid the significant influence 
of the high-temperature bright spot on the color.}

The $B-V$ color variation with respect to the $V$-band 
magnitude is displayed in the left panel of Figure \ref{color}.  
As described in \citet{ham20modeling}, the color of 
SS Cyg normally evolves counterclockwise with time 
through one outburst cycle in this color-magnitude 
diagram (see the arrows in the left panel), 
and the $B-V$ color in the normal quiescent state is 
around 0.7 (see the triangles).  
However, the $B-V$ color became bluer in Event A and 
much bluer most of the time in Event B (see 
the circles and the squares).  
The color during Event A and Event B seems to be similar to 
that in the fading stage from the outburst maximum.  
At the fading stage, the cooling wave terminating an outburst 
is triggered at the outer disk and propagates inwards 
\citep{min85DNDI}.  
The above-mentioned similarity implies that the inner region 
of the disk became hotter during Event A than in the normal 
quiescent state and that the hot inner region was heated up 
further and/or occupied a larger area in the disk during Event B.  
Additionally, strong Balmer emission lines and He II (4686) 
emission line were detected in the optical spectra taken 
in Event B (see Figure E3).  
In comparison with the time evolution of optical spectra 
in SS Cyg, which was reported by \citet{hes84sscyg}, 
these features are similar to those of optical spectra in 
the fading stage of the same system.  
We here note that all of the WD, the disk, 
the secondary star, and the bright spot contribute to 
the optical emission; however, the effect on the color variation 
by the change in the WD temperature is very small since the WD 
is tiny.  
Also, there is no observational evidence of a significant 
increase in the temperature of the secondary star.  
Moreover, we avoid the contamination of the bright spot.  

We also show the $V-I$ color variation with respect to 
the $V$-band magnitude in the right panel of Figure \ref{color}.  
The $V-I$ color evolves linearly in this diagram 
during normal long outbursts (see the triangles in the same panel).  
We see that this color became redder in Event A and Event B 
by $\sim$0.2 in comparison with the data in normal long 
outbursts.  
This may imply that the disk expanded in the radial direction 
in Event A and Event B.  
The typical disk spectrum is a multi-temperature blackbody 
and its Rayleigh-Jeans tail exists around optical and 
near-infrared wavelengths.  That tail depends on the temperature 
of the outermost disk and the disk radius.  
The $V-I$ color is more sensitive to them than the $B-V$ color.  
The redder $V-I$ color in Event A and Event B 
therefore may imply that the disk expanded in the radial direction 
and/or that the low-temperature component in the outer disk 
increased.  
The $V-I$ colors of 0.5 and 1.0 correspond to the blackbody 
radiation of $\sim$7,000~K and $\sim$5,000~K.

\subsection{Broad-band X-ray spectra}

\subsubsection{Model of multi-temperature plasma}

Non-magnetic CVs including DNe harbor X-ray emitting region 
in the vicinity of the WD, in which the rotational energy of 
the matter transferred from the inner edge of the accretion 
disk is deposited to enable the gas to settle onto 
the slowly-rotating WD.  
In the conventional picture, the X-ray emitting region named 
a boundary layer (BL) is optically thin with low accretion 
rates onto the WD, 
and becomes optically thick with high accretion rates 
and emits UV and soft X-ray photons 
\citep{nar93BL,pat85CVXrayemission1,pat85CVXrayemission2}.  
However, recent X-ray spectral and timing analyses suggest 
that optically-thin multi-temperature plasma emitting hard 
X-rays exists regardless of mass accretion rates.  
Although its structure and the mechanism for generating 
that region are controversial, an optically-thin hot accretion 
flow like advection-dominated accretion flow (ADAF; 
\cite{nar94ADAF}) sometimes merged with the BL may surround 
the WD and the accretion disk would be truncated apart 
from the WD surface (see \cite{bal20xrayCV} for a review).  
We here treat broad-band X-ray spectra during Event A and 
Event B, when the accretion rate would not be very high, 
and simply call the X-ray emitting region the BL.  

We obtained five sets of broad-band X-ray spectra in 
the 0.5--50~keV thanks to the coordinated observations by 
{\it NICER} and {\it NuSTAR} (see also the arrows 
in the upper panel of Figure \ref{multi-overall}) and 
attempt to fit these spectral energy densities (SEDs)\footnote{The {\it NuSTAR} spectrum extends to 80~keV, but we utilize the data in the 3--50~keV because the intrinsic spectrum from SS Cyg is comparable with the background in the $>$50~keV.  Also, the model that we use is limited to the energy band softer than $\sim$50~keV.  We confirm that there is no hard energy tail in these spectra.}.  
The broad-band spectra referred to as T1 and T2 in 
Figure \ref{broad-xray-spectra} were taken in 
Event A and those referred to as T3, T4, and T5 
were taken in Event B, respectively.  

In evaluating the X-ray spectrum of non-magnetic CVs, 
the cooling flow model is practically adopted 
(e.g., \cite{don97sscygXray,ish09sscygSuzaku,tsu18gammaCas,nak19vwhyi}).  
In this paper, we use the \texttt{CEVMKL} model in 
the software \texttt{XSPEC} \citep{XSPEC} 
for comparison with the spectral analyses of SS Cyg 
by \citet{ish09sscygSuzaku}.  
In this model, the expected spectrum is a sum of 
bremsstrahlung emissivity as follows: 
\begin{equation}
L(\nu) \propto \int^{T_{\rm max}}_{T_{\rm bb}} \frac{\epsilon (T, n^2, \nu)}{\epsilon(T, n^2)} dT. 
\label{emissivity}
\end{equation}
Here $\epsilon(T, n^2, \nu)$ is the bremsstrahlung spectral 
emissivity, $\epsilon(T, n^2)$ is the total bremsstrahlung 
emissivity integrated over $\nu$, 
and $T_{\rm max}$ and $T_{\rm bb}$ are the shock temperature 
at the boundary between the disk and the BL and the blackbody 
temperature of the WD surface and/or the surface of 
the inner disk, respectively.  
The differential emission measure (DEM) in this model is 
proportional to a power-law function of temperature 
\citep{don95eferi} and given as 
\begin{equation}
d{\rm (EM)} \propto (T/T_{\rm max})^{\alpha}~{d}(\log T) \propto (T/T_{\rm max})^{\alpha-1}~{d}T~~{\rm for}~T < T_{\rm max}.
\label{temp-powerlaw}
\end{equation}

The \texttt{CEVMKL} model is necessary to be combined by 
the reflection model because the surface of the WD and/or 
the accretion disk subtends the BL, intercepts X-rays, 
and reflects some of them off into space \citep{don97sscygXray}.  
Also, the \texttt{GAUSSIAN} model should be introduced 
to describe the fluorescence K$\alpha$ line emission.  
We therefore fit the SEDs with the model 
\texttt{TBABS*(REFLECT*CEVMKL + GAUSSIAN)}.  
We utilize the software \texttt{XSPEC} version 12.11.0.  
Here the \texttt{TBABS} model is multiplied in order to 
account for the interstellar and possible intrinsic absorption 
of SS Cyg.

\subsubsection{Results of spectral modeling}

As discussed by \citet{ish09sscygSuzaku}, the abundances 
of iron and oxygen in the \texttt{CEVMKL} model, 
which are denoted as $Z_{\rm Fe}$ and $Z_{\rm O}$, 
affect very much the determination of the solid angle of 
the reflector subtending the BL, which is denoted 
as $\Omega / (2\pi)$, and the plasma emission parameters 
$T_{\rm max}$ and $\alpha$ in equation (\ref{temp-powerlaw}).  
These abundances should be the same over the five datasets.  
Before analyzing each individual spectrum, we simultaneously 
fit all five datasets, tying $Z_{\rm Fe}$ and $Z_{\rm O}$ 
in the \texttt{CEVMKL} model.  
The abundances of Ca, Ar, S, Si, Mg, and Ne are fixed 
at the estimates by \citet{ish09sscygSuzaku}.  
The abundances of the other elements are fixed at 
solar values given in \citet{and89abundance}.  
The abundance and the iron abundance in the \texttt{REFLECT} 
model are tied to $Z_{\rm O}$ and $Z_{\rm Fe}$, 
respectively.  
The line energy in the \texttt{GAUSSIAN} model is 
fixed at 6.4~keV.  
The hydrogen column density ($N_{\rm H}$) in the \texttt{Tbabs} 
model, $\Omega / (2\pi)$ in the \texttt{REFLECT} model, 
$T_{\rm max}$, $\alpha$, and the normalization in 
the \texttt{CEVMKL} model, and the line width ($\sigma$) 
and the normalization in the \texttt{GAUSSIAN} model are 
free to vary.  
The inclination angle of the reflector is set to be 
that of the binary system, 45~deg \citep{hil17sscyg}.  
Here we use 114.25~pc as the distance to SS Cyg, which is 
obtained by the {\it Gaia} parallax \citep{bai18gaia}.  
Besides, we add an offset in normalizations between 
the {\it NICER} and {\it NuSTAR} SEDs in our fittings 
by considering uncertainties on the cross calibration and 
time variations of the emission from this source.  
The observational times by {\it NICER} and {\it NuSTAR} were 
not exactly the same in each coordinated observation and 
the X-ray flux of this source varies on timescales of 
hundreds of seconds.  
\textcolor{black}{
We also add a cross-normalization parameter between 
the {\it NuSTAR} FPMA and FPMB SEDs in T3--T5.}  
As a result, the best-fitting values for $Z_{\rm Fe}$ and 
$Z_{\rm O}$ are 0.10 and 0.23, respectively.  
The normalization of the {\it NICER} SEDs 
is 0.83--1.02 times of that of the {\it NuSTAR} SEDs.  
\textcolor{black}{
The normalization of the {\it NuSTAR} FPMB is 
1.01--1.05 times of that of the {\it NuSTAR} FPMA 
in T3--T5.}  

\begin{figure*}[htb]
\vspace{-5mm}
\begin{center}
\begin{minipage}{0.49\hsize}
\FigureFile(80mm, 50mm){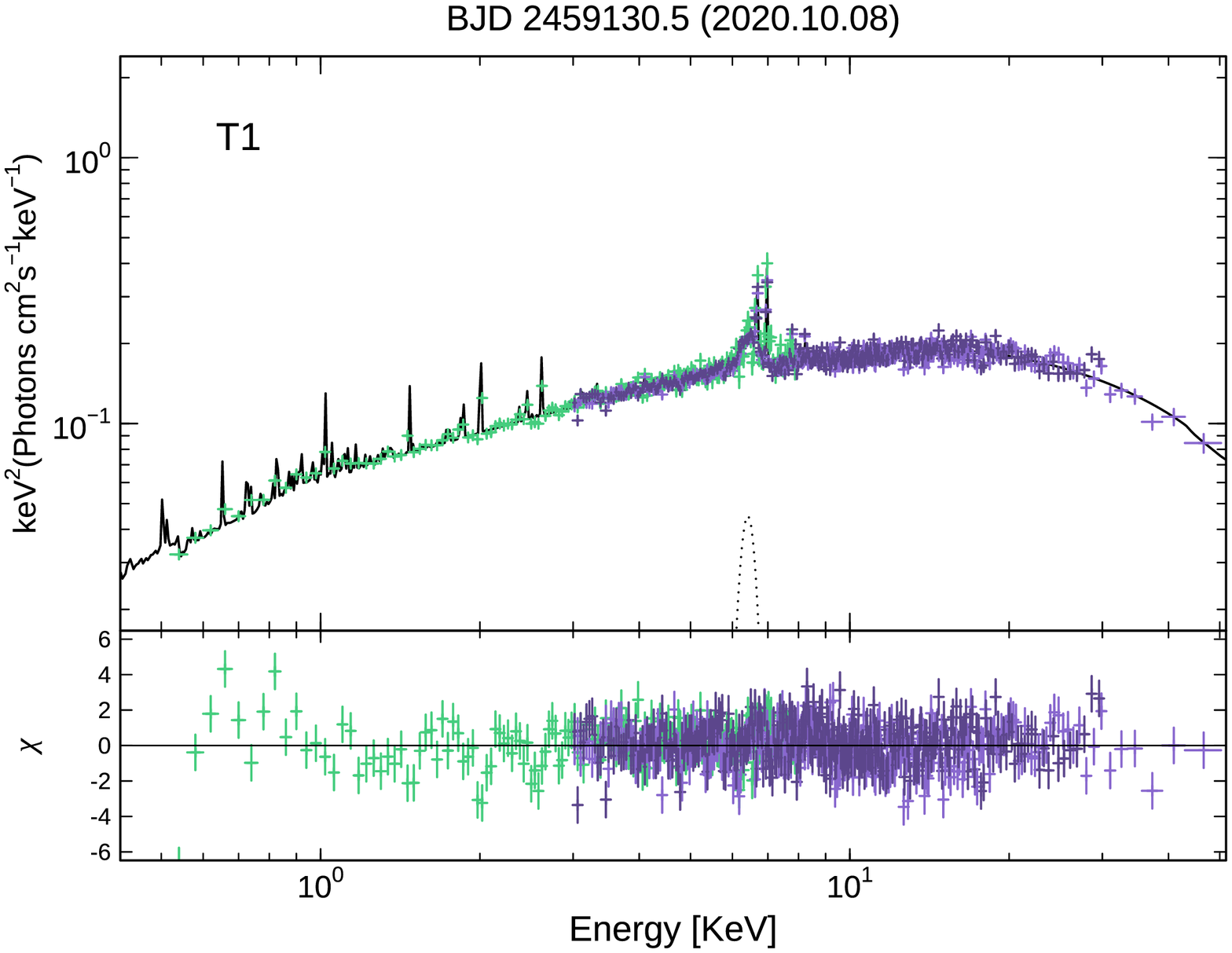}
\end{minipage}
\begin{minipage}{0.49\hsize}
\FigureFile(80mm, 50mm){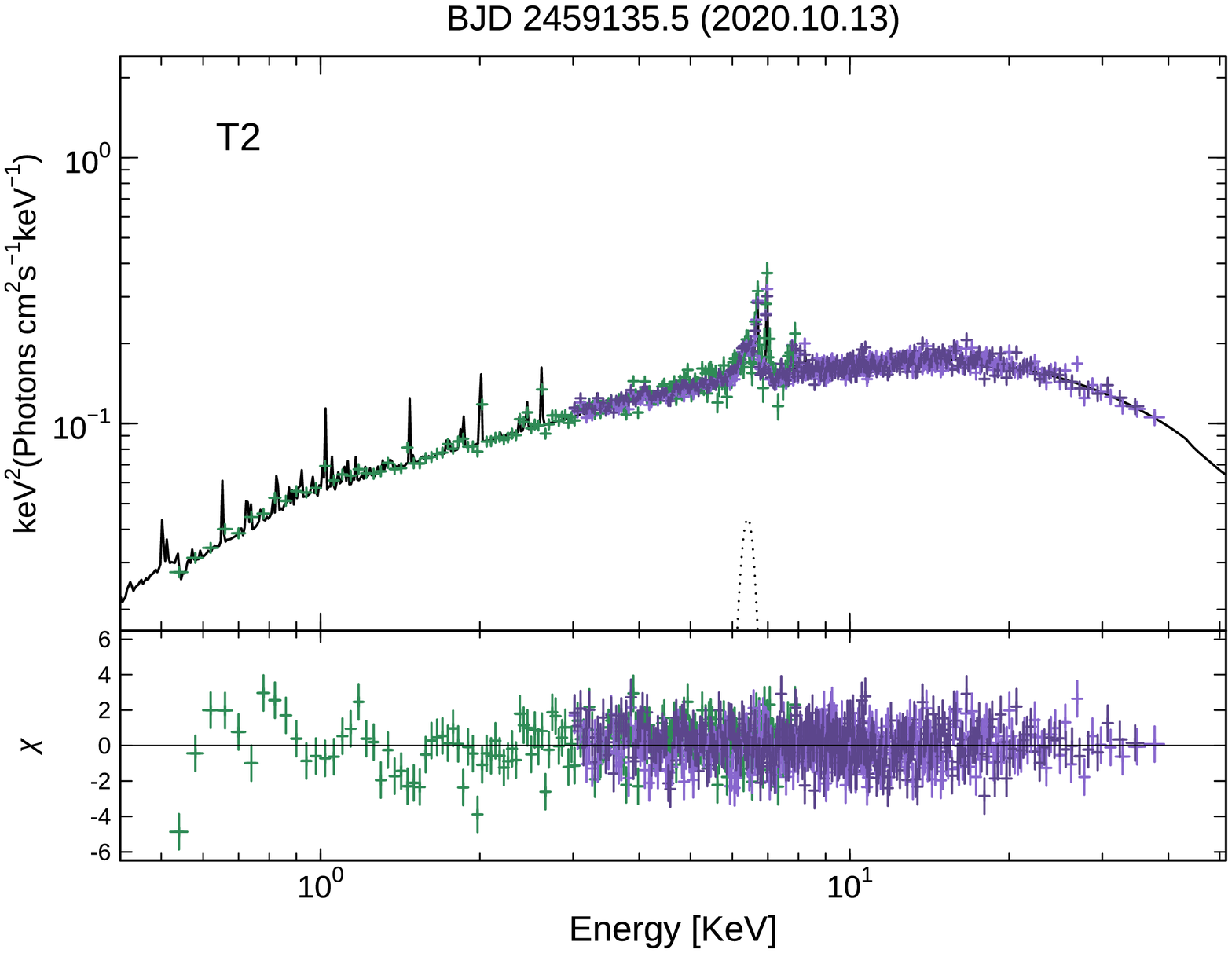}
\end{minipage}
\\
\vspace{-3mm}
\begin{minipage}{0.49\hsize}
\FigureFile(80mm, 50mm){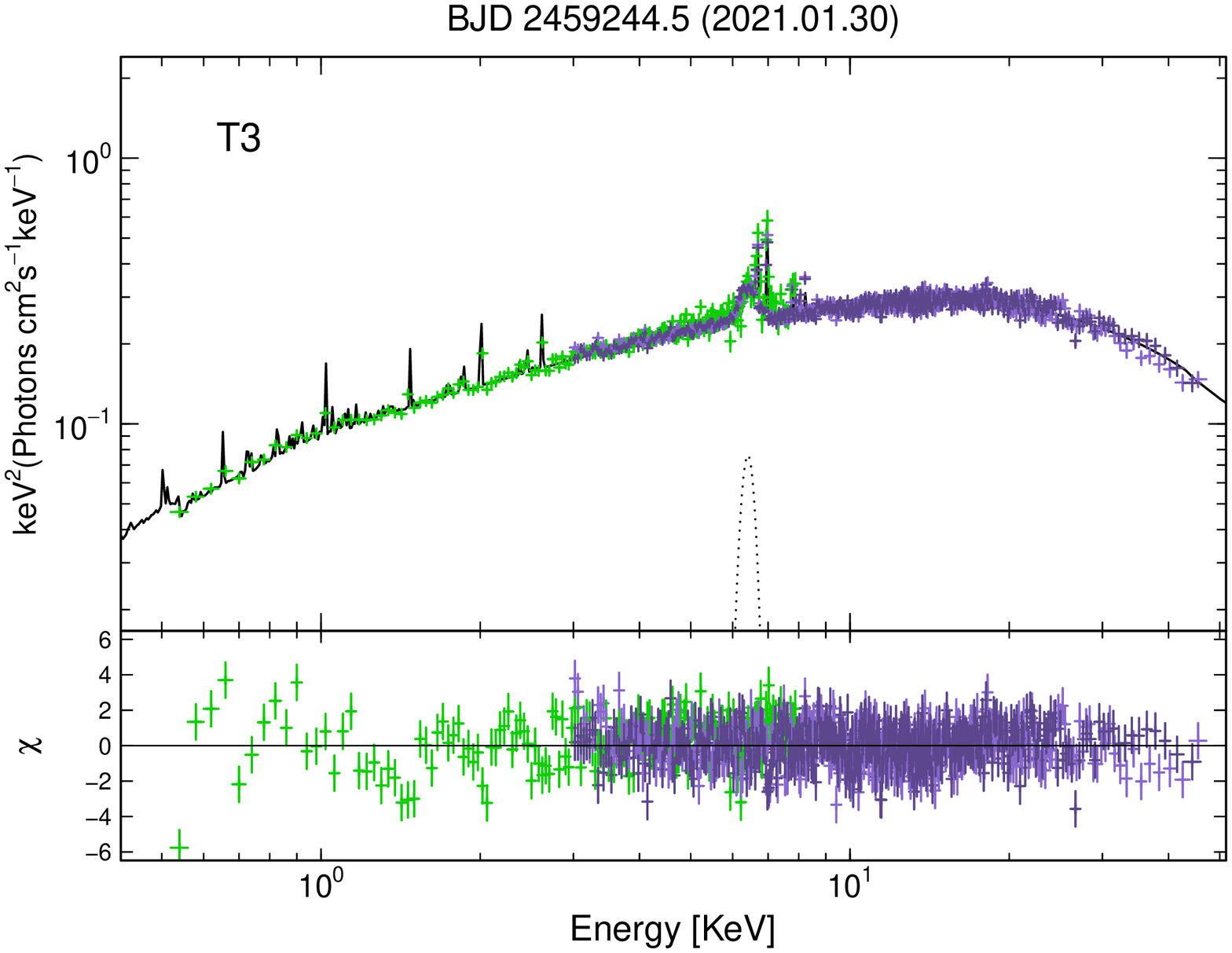}
\end{minipage}
\begin{minipage}{0.49\hsize}
\FigureFile(80mm, 50mm){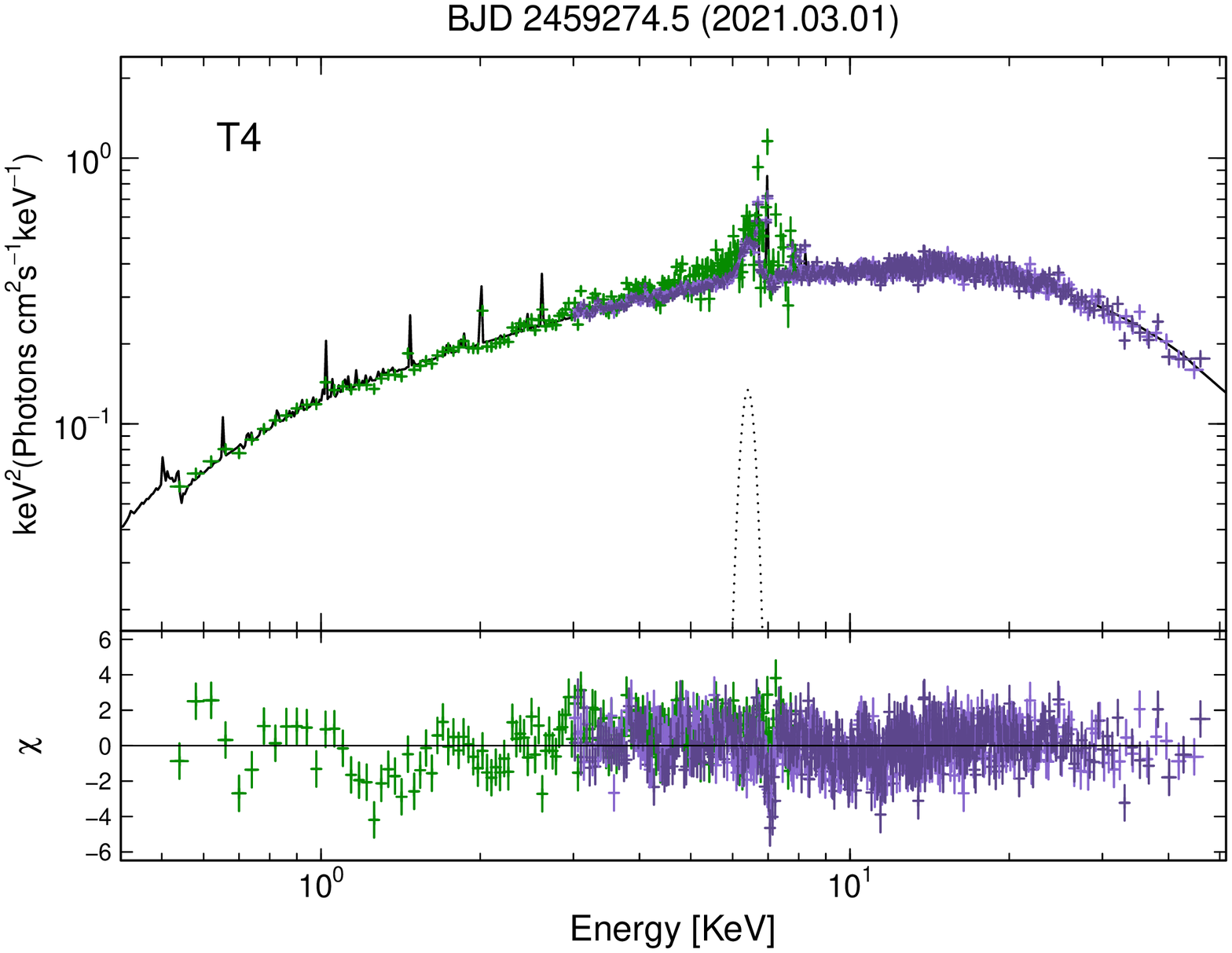}
\end{minipage}
\\
\vspace{-3mm}
\begin{minipage}{0.49\hsize}
\FigureFile(80mm, 50mm){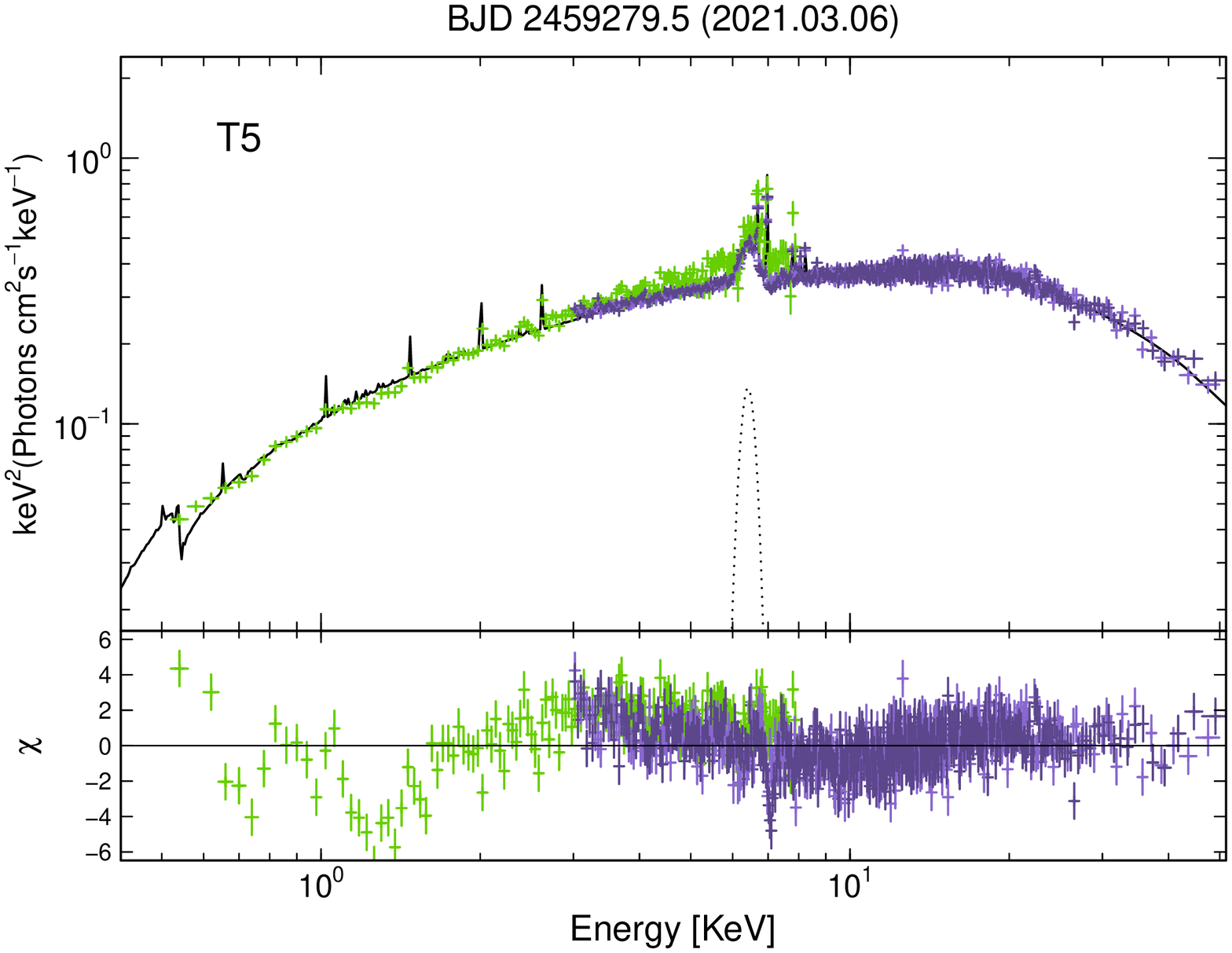}
\end{minipage}
\begin{minipage}{0.49\hsize}
\FigureFile(80mm, 50mm){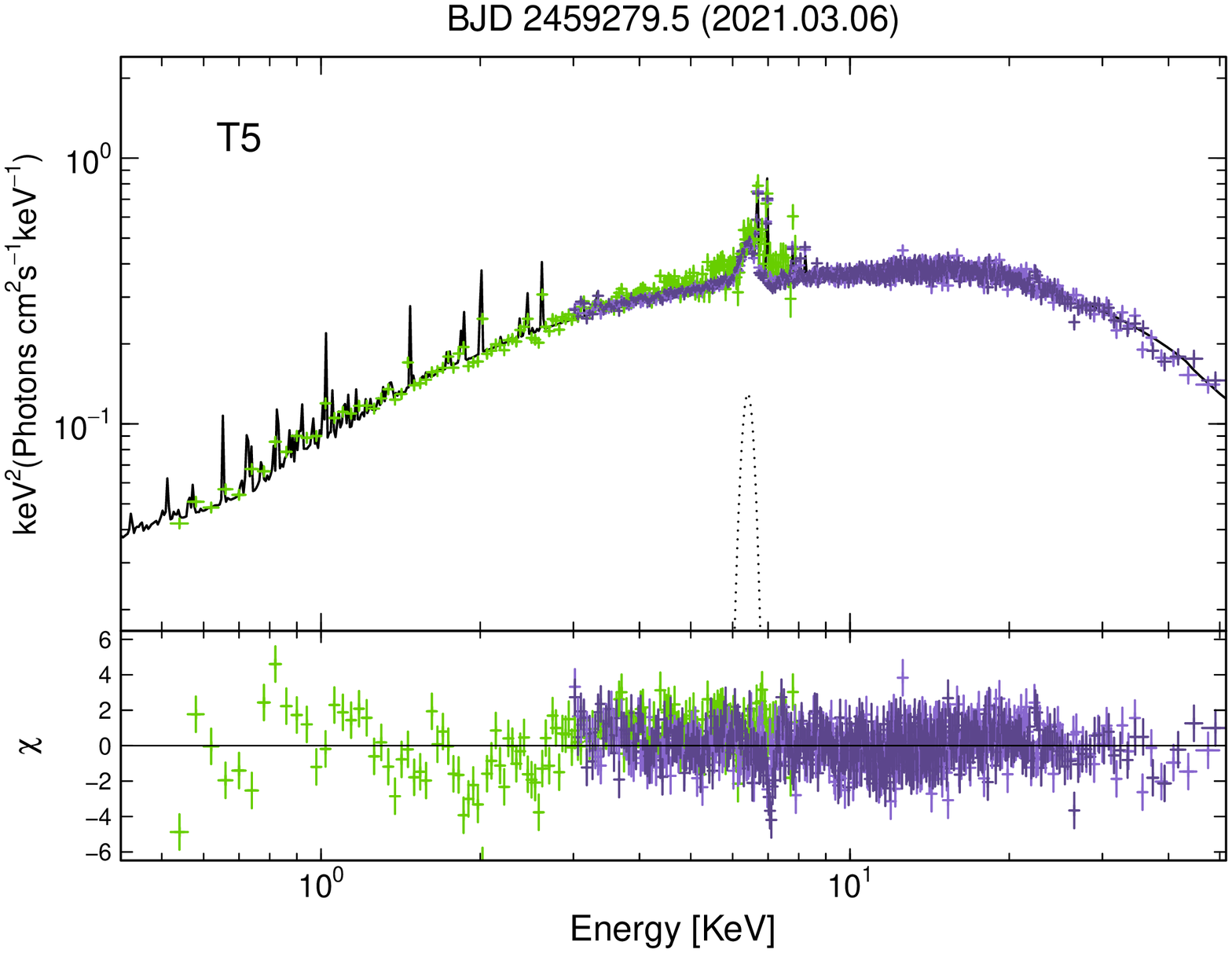}
\end{minipage}
\end{center}
\caption{Broad-band X-ray SEDs.  The green crosses represent the observed {\it NICER} SEDs.  The purple and dark purple crosses represent the observed {\it NuSTAR} FPMA and FPMB SEDs, respectively.  The solid and the dot lines represent the emission reproduced by our model and the model emission from the fluorescence K$\alpha$ line, respectively.  }
\label{broad-xray-spectra}
\end{figure*}

\begin{table*}[htb]
    \caption{Best-fitting parameters of our modeling in the five sets of broad-band spectra.  The errors represent 90\% confidence ranges.  }
    \vspace{2mm}
    \label{parameter-cevmkl}
    \centering
    \begin{tabular}{cccccccc}
\hline
Model & Parameter & T1 & T2 & T3 & T4 & T5 & T5 \\
\hline
Tbabs & $N_{\rm H}$$^{*}$ 
      & 2.7$\pm$0.2 
      & 3.6$\pm$0.2 
      & 2.8$\pm$0.2 
      & 4.9$\pm$0.3 
      & 8.6$\pm$0.1 
      & -- \\
\hline
pcfabs & $N_{\rm H}$$^{*}$ 
      & -- 
      & -- 
      & -- 
      & -- 
      & -- 
      & 70.0$\pm$3.3 \\
      &
      $f$$^{\dagger}$ 
      & -- 
      & -- 
      & -- 
      & -- 
      & -- 
      & 0.59$\pm$0.01 \\
\hline
reflection & $\Omega / (2\pi)$$^{\ddagger}$ 
        & 0.27$\pm$0.04
        & 0.28$^{+0.05}_{-0.04}$ 
        & 0.23$\pm$0.03 
        & 0.15$\pm$0.03 
        & 0.10$\pm$0.02 
        & 0.25$\pm$0.03 \\
\hline
cevmkl & $\alpha$$^{\S}$ 
        & 0.82$\pm$0.03 
        & 0.85$\pm$0.03 
        & 0.96$\pm$0.03 
        & 1.25$^{+0.07}_{-0.06}$ 
        & 1.86$\pm$0.10 
        & 0.65$\pm$0.04 \\
        & $T_{\rm max}$$^{\P}$ 
        & 33.8$\pm$0.7 
        & 32.6$^{+1.2}_{-1.1}$ 
        & 33.7$\pm$0.7 
        & 27.1$\pm$0.5 
        & 23.6$\pm$0.4 
        & 30.7$\pm$0.6 \\
        & N$_1$$^{|}$ 
        & 0.52$\pm$0.01 
        & 0.49$\pm$0.01 
        & 0.88$\pm$0.02 
        & 1.56$\pm$0.06 
        & 2.06$\pm$0.08 
        & 1.04$^{+0.04}_{-0.03}$ \\
\hline
gaussian & $\sigma$$^{**}$ 
        & 0.22$\pm$0.04 
        & 0.20$\pm$0.03 
        & 0.20$\pm$0.02 
        & 0.20$\pm$0.01 
        & 0.21$\pm$0.01 
        & 0.18$\pm$0.02 \\
        & N$_2$$^{\dagger\dagger}$ 
        & 6.06$^{+0.63}_{-0.59}$ 
        & 5.50$^{+0.54}_{-0.52}$ 
        & 9.31$^{+0.54}_{-0.53}$ 
        & 16.8$^{+0.77}_{-0.76}$ 
        & 17.9$\pm$0.49 
        & 14.2$^{+0.75}_{-0.74}$ \\
\hline
$\chi^2$/dof & & 1.30 & 1.22 & 1.36 & 1.40 & 2.12 & 1.44 \\
\hline
$L_{\rm X}$$^{\S\S}$ & 
        & 1.4$\times$10$^{33}$ 
        & 1.2$\times$10$^{33}$ 
        & 2.1$\times$10$^{33}$ 
        & 2.8$\times$10$^{33}$ 
        & 2.7$\times$10$^{33}$ 
        & 2.7$\times$10$^{33}$ \\
\hline
Hardness$^{\ddagger\ddagger}$ & & 1.21 & 1.21 & 1.23 & 1.19 & 1.21 & 1.18 \\
\hline
\multicolumn{8}{l}{$^{*}$Equivalent hydrogen column in units of 10$^{20}$ atoms cm$^{-2}$.}\\
\multicolumn{8}{l}{$^{\dagger}$Dimensionless covering fraction ($0 < f \leq 1$).}\\
\multicolumn{8}{l}{$^{\ddagger}$Solid angle of cold matter surrounding the plasma.}\\
\multicolumn{8}{l}{$^{\S}$Index for power-law emissivity function.}\\
\multicolumn{8}{l}{$^{\P}$Maximum temperature in units of keV.}\\
\multicolumn{8}{l}{$^{|}$Normalization of the plasma model.  
}\\
\multicolumn{8}{l}{$^{**}$Line width in units of keV.}\\
\multicolumn{8}{l}{$^{\dagger\dagger}$Total photons/cm$^2$/s in the line of sight in units of 10$^{-4}$.}\\
\multicolumn{8}{l}{\parbox{400pt}{$^{\ddagger\ddagger}$X-ray luminosity in the 1--50-keV band in units of erg~s$^{-1}$.  
}}\\
\multicolumn{8}{l}{$^{\S\S}$Hardness ratio defined as the ratio of the flux in the 6--15 keV to that in the 2--6 keV.}\\
\end{tabular}
\end{table*}

We next fit each individual SED by the same model 
with $Z_{\rm Fe}$ and $Z_{\rm O}$ fixed at the obtained 
values.  
The extracted best-fitting parameter values and their errors 
for each SED are summarized in Table \ref{parameter-cevmkl} 
and the SEDs and the models are displayed 
in Figure \ref{broad-xray-spectra}.  
We also give the X-ray luminosity and the hardness ratio 
in the same table.  
Our modeling of these SEDs works well for the data of T1--T4.  
We see some excess for the SED in T5 in the softer energy 
band less than 8~keV and the fit is improved if we include 
the \texttt{pcfabs} model for partial absorption instead of 
the \texttt{Tbabs} model (see the bottom panels of 
Figure \ref{broad-xray-spectra}).  
The results are added in Table \ref{parameter-cevmkl}.  

According to the best-fitting parameters given 
in Table \ref{parameter-cevmkl}, 
the solid angle subtended by the reflector is small for all SEDs.  
The maximum temperature of the BL exceeds 30~keV 
in T1--T3 and decreases in T4 and T5.  
The hardness ratio defined as the ratio of the flux in 
the 2--6 keV with respect to that in the 6--15 keV is 
almost the same from T1 to T5.  
In comparison with the estimates by \citet{whe03sscyg} and 
\citet{ish09sscygSuzaku}, 
the X-ray luminosity increases by about twice in T1 and T2, 
by $\sim$4 times in T3, and by $\sim$5 times in T4 and T5.  
The width of the fluorescence K$\alpha$ line is obtained 
as $\sim$0.2~keV in all of the observations, though 
this is further investigated in sub-subsection 3.4.3 
by using only the narrow energy range of the {\it NICER} 
data to avoid systematic uncertainties on 
the cross calibration and the simultaneous fitting of 
wide-band spectra.  

The derived hydrogen column density of photoelectric 
absorption is much larger than the interstellar absorption 
of SS Cyg, 3.0$\times$10$^{19}$~cm$^{-2}$, 
which was constrained by ultraviolet and optical spectroscopy 
\citep{mau88CVinterstellarIUE,rit13sscyginterstellar}.  
This implies that additional material responsible for 
intrinsic absorption surrounds the BL.  
This absorption may change on timescales of around 
one week according to the results of SED fittings in 
T1 and T2 and those in T4 and T5 and seems to become 
larger from T3 to T5.  
The improvement of the SED fitting in T5 with the model 
for partial absorption may suggest that the BL in T5 is 
partially masked by dense absorbers.  
Although the ionization degree of the absorber might change 
between T1--T5, a detailed study is deferred to 
our subsequent paper.

\subsubsection{Iron emission lines}

\begin{figure}[htb]
\begin{center}
\FigureFile(80mm, 50mm){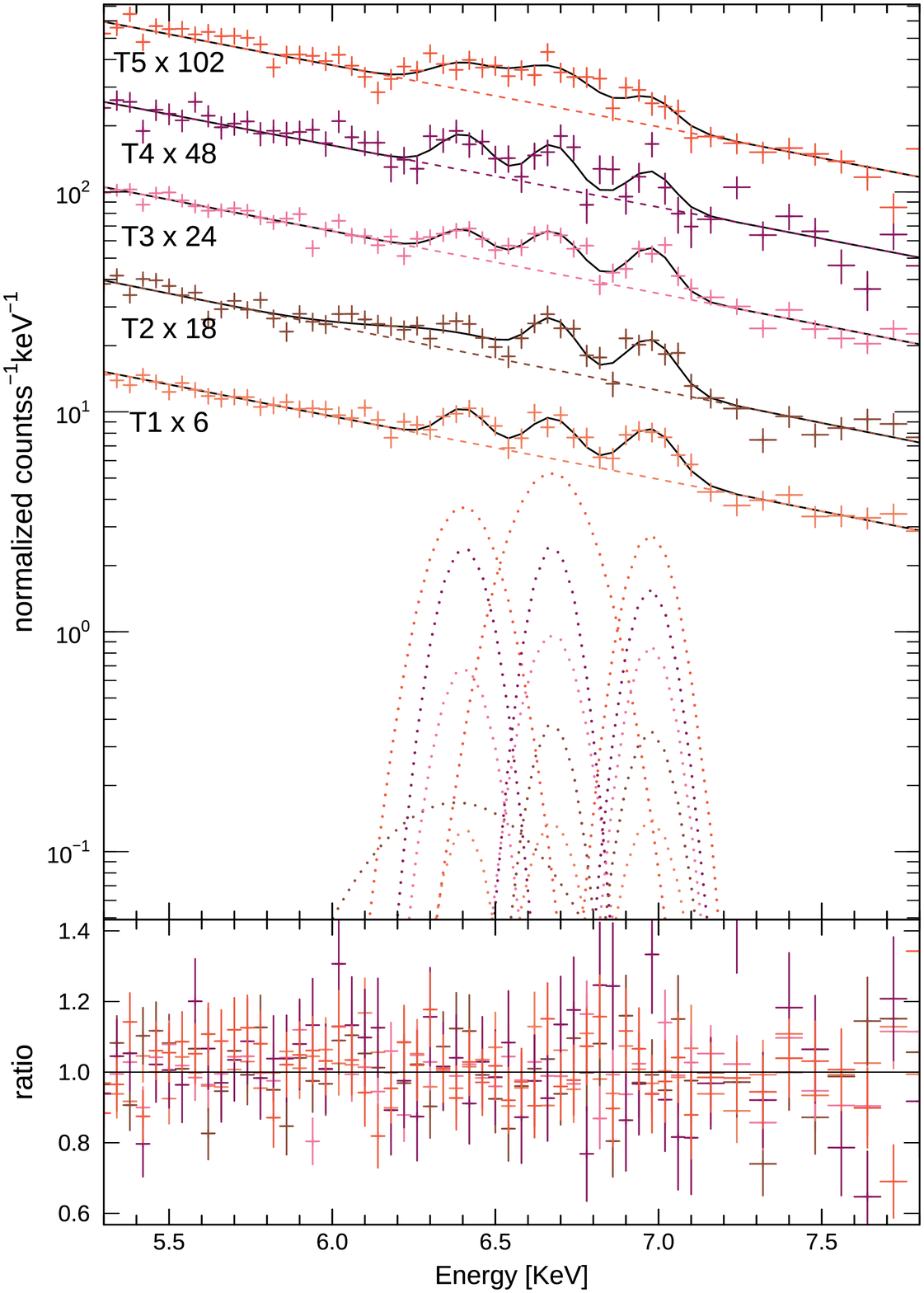}
\end{center}
\caption{Results of the model fitting by a power law plus three Gaussians.  The solid, dashed, and dot lines represent the emission reproduced by our model, the model continuum, and the model emissions from the three iron lines, respectively.  Each observed spectrum and each continuum model are shifted in the vertical axis for visibility and their offsets are denoted at the left side of each spectrum.  We set another offset for the three Gaussian models of each spectrum for visibility.  
}
\label{iron-lines}
\end{figure}

We next attempt to fit the {\it NICER} spectra in 
the 5--8~keV by a model consisting of a power-law 
continuum and three Gaussian lines in order to 
evaluate the fluorescence, He-like, and H-like K$\alpha$ 
iron emission lines.  
The central energy of the three Gaussian lines is fixed 
at 6.4, 6.67, and 6.97~keV.  
We set the lower limit of the line width at 1~eV 
because the energy resolution of the Silicon Drift 
Detector (SDD) on board {\it NICER} is 137~eV at 
6~keV\footnote{$<$https://heasarc.gsfc.nasa.gov/docs/nicer/nicer\_tech\_desc.html$>$}.  
The fitting results are shown in Figure \ref{iron-lines}.  
The best-fitting parameter values are given in 
Table \ref{parameter-lines}.  

\begin{table*}[htb]
    \caption{Best-fitting parameters of our modeling in the five sets of the {\it NICER} spectrum in the 5--8 keV.  The errors represent 90\% confidence ranges.  }
    \vspace{2mm}
    \label{parameter-lines}
    \centering
    \begin{tabular}{ccccccc}
\hline
Model & Parameter & T1 & T2 & T3 & T4 & T5\\
\hline
power law (continuum) & $\alpha$$^{*}$
        & 1.6$\pm$0.2 
        & 1.7$\pm$0.2 
        & 1.6$\pm$0.1 
        & 1.5$\pm$0.1 
        & 1.5$\pm$0.2 \\
        & $N_{\rm c}$$^{\dagger}$
        & 7.6$^{+2.7}_{-1.9}$ 
        & 7.7$^{+1.5}_{-1.2}$ 
        & 12.1$^{+3.4}_{-2.6}$ 
        & 14.0$\pm$3.2 
        & 15.0$^{+5.8}_{-4.0}$ \\
\hline
gaussian 1 (fluorescence) 
        & $\sigma_1$$^{\ddagger}$ 
        & 0.016$^{+0.057}_{-0.016}$ 
        & 0.236$^{+0.092}_{-0.064}$
        & 0.040$^{+0.048}_{-0.040}$ 
        & 0.038$^{+0.047}_{-0.038}$ 
        & 0.078$^{+0.067}_{-0.056}$ \\
        & N$_1$$^{\S}$
        & 2.7$^{+1.0}_{-0.8}$ 
        & 4.4$\pm$1.4 
        & 4.1$^{+1.5}_{-1.3}$ 
        & 7.1$\pm$2.4 
        & 7.0$^{+3.5}_{-4.9}$ \\

        & EW$_1^{\P}$
        & 60$^{+24}_{-17}$ 
        & 115$^{+85}_{-50}$ 
        & 53$^{+21}_{-19}$ 
        & 74$^{+45}_{-22}$ 
        & 65$^{+31}_{-30}$ \\
\hline
gaussian 2 (He-like) & 
        $\sigma_2$$^{\ddagger}$
        & 0.032$^{+0.044}_{-0.032}$ 
        & 0.001$^{+0.042}_{-0.001}$ 
        & 0.040$^{+0.030}_{-0.040}$ 
        & 0.001$^{+0.064}_{-0.001}$ 
        & 0.094$^{+0.117}_{-0.050}$ \\
        & N$_2$$^{\S}$
        & 3.5$\pm$1.0 
        & 2.9$\pm$0.7 
        & 6.6$\pm$1.5 
        & 7.0$\pm$1.7 
        & 12.7$^{+9.4}_{-3.9}$ \\

        & EW$_2^{\P}$
        & 76$^{+32}_{-22}$ 
        & 74$\pm$25 
        & 85$^{+22}_{-19}$ 
        & 69$^{+34}_{-25}$ 
        & 127$^{+52}_{-40}$ \\
\hline
gaussian 3 (H-like) & 
        $\sigma_3$$^{\ddagger}$ 
        & 0.036$^{+0.035}_{-0.036}$
        & 0.027$^{+0.036}_{-0.027}$
        & 0.001$^{+0.051}_{-0.001}$ 
        & 0.001$^{+0.070}_{-0.001}$ 
        & 0.034$^{+0.095}_{-0.033}$ \\
        & N$_3$$^{\S}$ 
        & 4.2$\pm$1.0 
        & 3.4$\pm$0.8 
        & 5.8$\pm$1.2 
        & 5.2$\pm$2.4 
        & 4.8$\pm$3.8 \\

        & EW$_3^{\P}$
        & 109$^{+33}_{-24}$ 
        & 98$^{+35}_{-33}$ 
        & 84$^{+19}_{-16}$ 
        & 49$^{+33}_{-22}$ 
        & 52$^{+43}_{-34}$ \\
\hline
$\chi^2$/dof & & 0.48 & 1.06 & 0.93 & 1.12 & 1.05 \\
\hline
\multicolumn{7}{l}{$^{*}$Photon index of power law $\alpha$.}\\
\multicolumn{7}{l}{$^{\dagger}$Total photons/cm$^2$/s in the line of sight of the continuum emission in units of 10$^{-2}$.}\\
\multicolumn{7}{l}{$^{\ddagger}$Line width in units of keV.}\\
\multicolumn{7}{l}{$^{\S}$Total photons/cm$^2$/s in the line of sight in units of 10$^{-4}$.}\\
\multicolumn{7}{l}{$^{\P}$Equivalent width in units of eV.}\\
\end{tabular}
\end{table*}

The width of the fluorescence line becomes broader 
in T5, although it is not well determined in T2.  
Part of the BL is considered to cover the innermost region 
of the disk.  
If the 6.4~keV emission line originates from the reflection 
at the innermost disk and if its line width 
is determined by the Keplerian velocity amplitude of 
the disk, 
the broader line width in T5 may imply that the inner disk 
edge gradually contracts just before T5.  
For example, the line width of 0.078$^{+0.067}_{-0.056}$~keV 
in T5 corresponds to the inner disk edge of 
(0.3--10)$\times$10$^{9}$~cm, which is roughly consistent with 
the estimation in subsection 3.5.  
The width of the He-like line seems to be broader in T5 
than those in the other time periods despite large 
error bars.  
The K$\alpha$ photons reflected by the WD and/or the disk 
may be suffered from Compton scattering as discussed in 
\citet{hel98CVlinewidth}.  
Some K$\alpha$ emission lines other than the three lines 
that we here consider might be mixed.  
Besides, the H-like emission line seems to become weaker 
from T3 to T4.  
This may be consistent with that the maximum temperature 
of the BL becomes lower in T4 and T5 in comparison with 
that in T1--T3 (see also Table \ref{parameter-cevmkl}) 
since the flux ratio between H-like and He-like emission 
lines is sensitive to the temperature of the X-ray 
emitting plasma in CVs and is expected to increase 
with increasing plasma temperature \citep{xu19ironlines}.

\subsection{X-ray stochastic variability}

Multi-wavelength light curves of some of CVs are 
more or less dominated by stochastic variability 
referred to as ``flickering'' 
(see \cite{bru92CVflickering,bru21flikering} for reviews).  
The flickering in CVs, X-ray binaries, and active galactic 
nuclei is considered to be produced by fluctuations of 
mass accretion rates at different disk radii, which vary 
on local viscous timescales at each radius of the disk 
\citep{lyn97flickering,yon97CVflickering,utt01flickering,are06flickering}.  
The fluctuations propagate inwards and characterize 
the X-ray variability from the BL.  
It is known that the power spectrum (PS) of 
flickering has a broken power-law shape and 
the break frequency is considered to be associated 
with the dynamical timescale at the inner disk edge 
(e.g., \cite{rev10IPpowerspec,bal12xrayDNe}), 
though other complex factors are related to it, 
as pointed out by \citet{sca14flickering}.  
We therefore try to estimate the innermost radius of 
the accretion disk by using the break frequency of the PS 
of X-ray light curves.  

We have extracted light curves in the 0.3--7-keV band 
with 1 s bins during Event A and Event B from 
the {\it NICER} data and divided the light curve into 
several windows in which the data have equally spaced 
sampling times.  
We have adopted the publicly available \texttt{Python} library 
\texttt{Stingray} \citep{hup19stingray} to 
each window and obtained the PS.  
Here we normalize the power to squared fractional 
root mean square \citep{bel90aperiodic} and bin the power 
with a logarithmic rebinning factor of 0.25 or 0.4.  
We fit each PS by the following equation: 
\begin{equation}
P(\nu) = a \nu^{-1} \left(1 + \left(\frac{\nu}{b} \right)^4 \right)^{-1/4} + c, 
\label{psd-fitting}
\end{equation} 
where $\nu$ is the frequency and $P(\nu)$ is the power.  
The parameters $a$, $b$, and $c$ are free to vary 
in our fitting, and $a$ and $b$ represent the normalization 
and the break frequency, respectively.  
We introduce the parameter $c$ to describe the white-noise 
component.  
This equation is based on a simple analytical model 
of the truncated accretion disk \citep{rev10IPpowerspec} 
and here $b$ is $\sqrt{G M_1 / r_{\rm in}^3} / (2\pi)$, 
where $G$ is the gravitational constant, $M_1$ is the WD mass, 
and $r_{\rm in}$ is the innermost radius of the accretion disk.  

\begin{figure}[htb]
\begin{center}
\FigureFile(80mm, 50mm){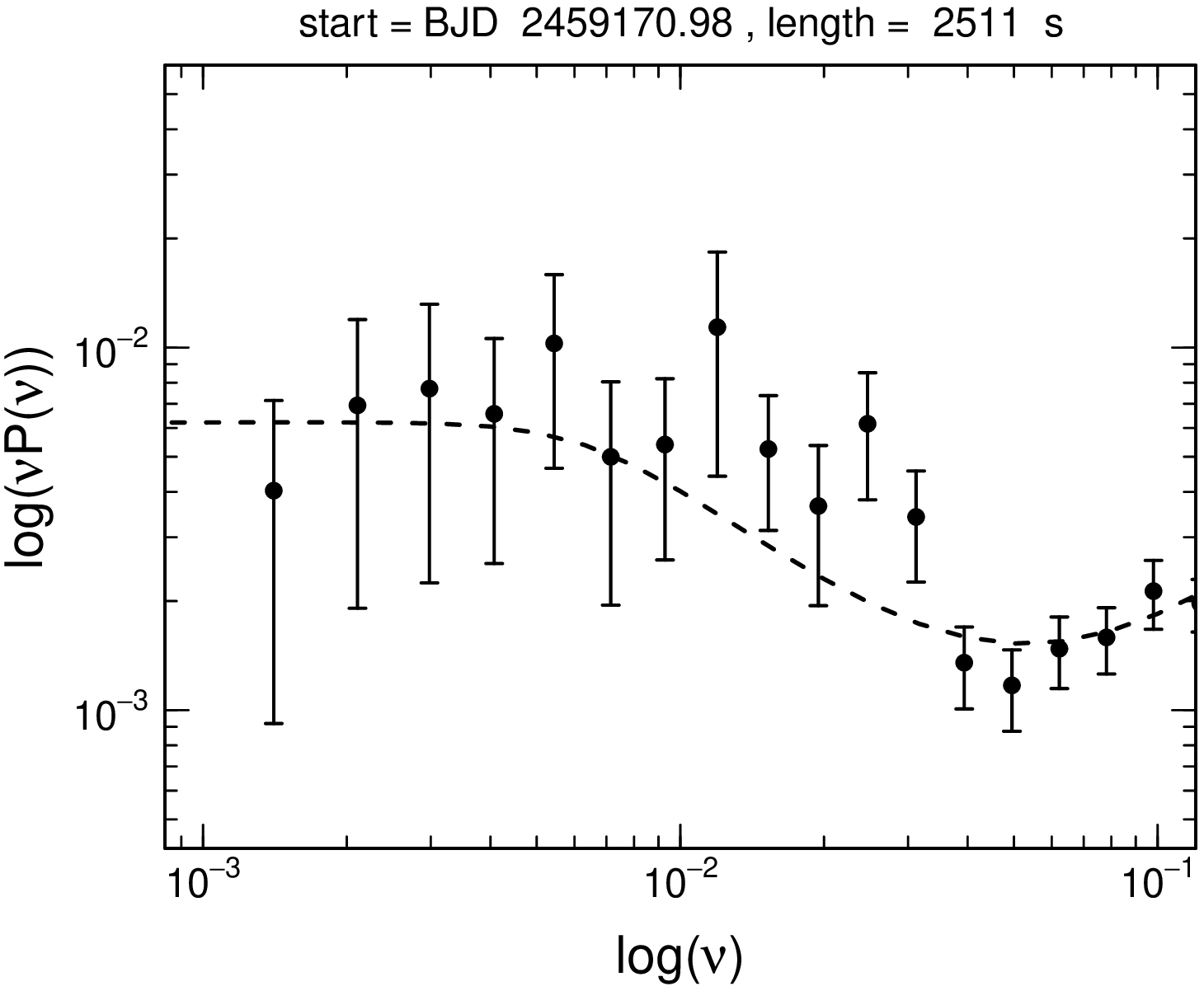}
\end{center}
\caption{
Power spectrum and fitting result of the {\it NICER} light curve during BJD 2459170.9752--2459171.0043.  
The black points and bars penetrating them represent the powers multiplied by the frequencies and their 1$\sigma$ errors.  The dashed line stands for the result of our fitting by using equation (\ref{psd-fitting}).  The start time of the observation and the length of the window are given at the top of this panel.  
}
\label{psd-example}
\end{figure}

The results of our fitting of PSs and the estimation of 
the innermost radius of the disk are tabulated in 
Table E4.  
An example of the PSs and their fitting results is 
displayed in Figure \ref{psd-example}, and 
the others are given in Figures E4 and E5.  
Although some light variations other than flickering may be 
excited on short timescales, we here simply assume that the PS 
has a power-law shape.  
The innermost disk radius is estimated to be 
(3.5$\pm$0.3)$\times$10$^9$~cm by averaging the results for 
each PS during Event A.  
Although this value is slightly smaller than the averaged value 
during the normal quiescence before the pre-stage, 
which is $\sim$5$\times$10$^9$~cm \citep{bal12xrayDNe}, 
our estimates are consistent with the idea that the disk is 
truncated far from the WD surface in the optical quiescence, 
which was proposed by \citet{mey94siphonflow}.  

\begin{figure}[htb]
\begin{center}
\FigureFile(80mm, 50mm){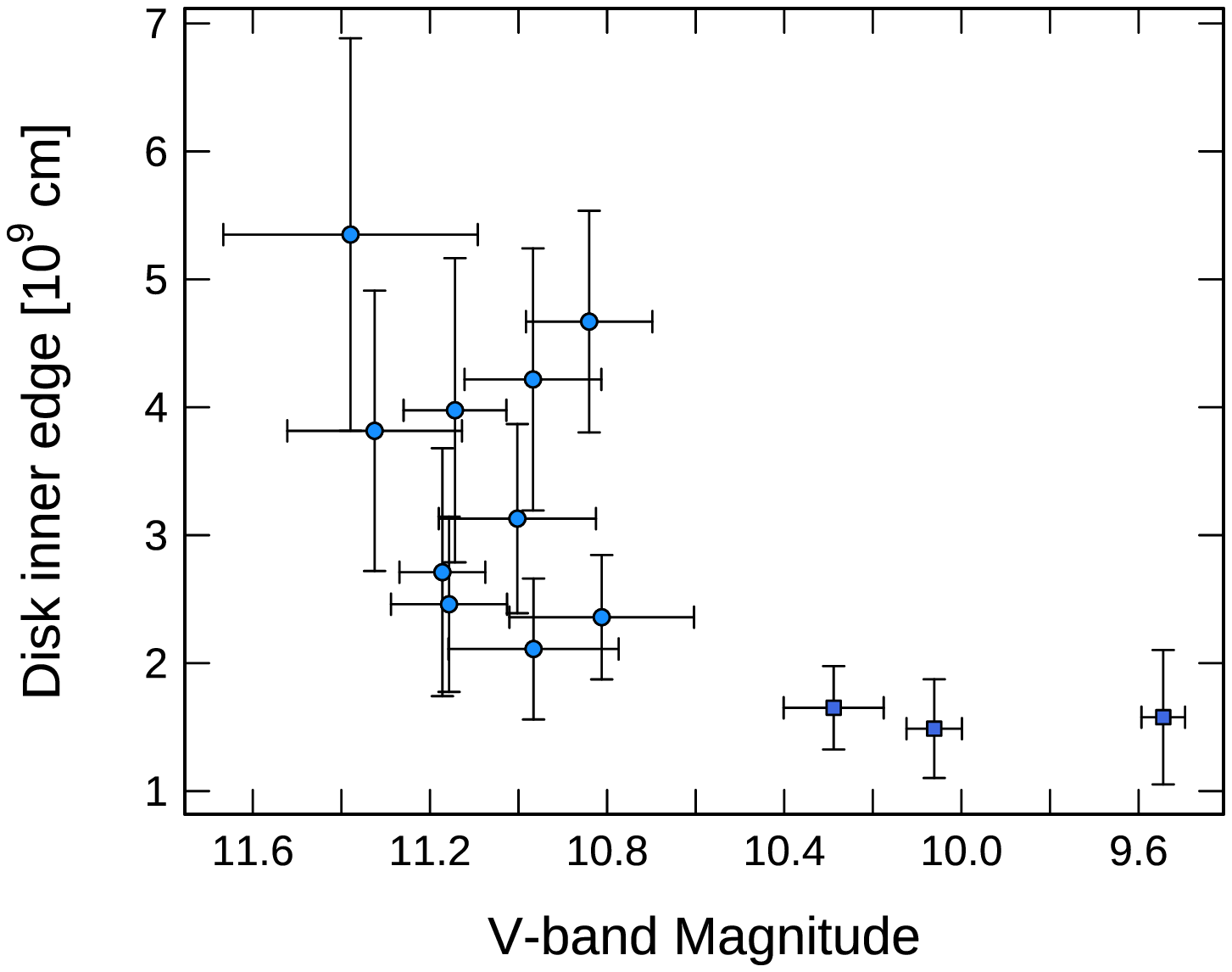}
\end{center}
\caption{
Correlation between the estimated size of the inner edge of the accretion disk and the optical $V$-band magnitude.  The circles and squares represent the estimates during Event A and those during Event B, respectively.  
}
\label{radius-vmag}
\end{figure}

It may be unclear whether the break frequency in PSs 
corresponds to the innermost disk radius or not; 
however, we yield the anti-correlation between the size of 
the inner edge of the disk and the optical brightness during 
Event A as displayed in Figure \ref{radius-vmag} despite 
large error bars (see the circles).  
This may imply that the inner disk edge extends toward the WD 
as the accretion rate of the disk increases.  
The WD radius is $\sim$5$\times$10$^{8}$~cm for the WD mass 
of 0.94$M_{\solar}$ \citep{nau72WDmassradius,pro98WDmassradius,hil17sscyg}.  
By contrast, the innermost disk radius appears to stay 
at a smaller value during Event B 
(see the squares in the same figure).

\section{Discussion}

\subsection{Cause of the simultaneous increase in the optical and X-ray flux}

Our ultimate goal is to understand why the anomalous 
event in 2021 in SS Cyg, i.e., Event B, was triggered.  
To achieve this purpose, it is essential to examine 
the cause of its multi-wavelength predecessor phenomenon 
which is the gradual increase in the optical flux 
in the quiescent state during the pre-stage, i.e., Event A.  
The following three scenarios should be discussed as 
its cause.  
The first possibility is an enhancement of mass 
transfer rates from the secondary star, which 
would also increase the X-ray flux after 
viscous timescales at the outer disk.  
The second possibility is that X-ray irradiation of 
the disk and the secondary star becomes stronger 
together with the increase in the X-ray flux.
The third possibility is that the intrinsic flux of 
the disk increases.  

\textcolor{black}{
First of all, we do not find any amplification of 
the orbital hump during Event A (see Figure \ref{rband-orbital}).  
The amplitude of the orbital hump in the flux scale 
should be enhanced if the luminosity of the bright spot, 
which is connected to the mass transfer rate, increases.  
There is no positive evidence that the mass transfer rate 
changes on average during Event A at least in our observations, 
although we cannot rule out that the transfer rate might 
fluctuate within the error bars, $\sim$5\%, on average and 
might temporarily vary.}  
Also, the irradiation of the secondary star that we discuss 
later cannot largely enhance the transfer rate 
\citep{osa04EMT}.  

We next examine X-ray irradiation of the disk and 
the secondary star.  
X-rays harder than $\sim$1~keV is able to heat the surface 
of the disk and the secondary star 
\citep{all59UVspec,cru74softXrayEUV}.  
\textcolor{black}{
The X-ray luminosity in the 1--50~keV band is estimated 
to be $\sim$1$\times$10$^{33}$~erg~s$^{-1}$ during Event A 
from our results in subsection 3.4 (see 
Table \ref{parameter-cevmkl}).  
The X-ray albedo of the disk is $\sim$0.9 according to 
\citet{dejon96LMXBreprocessing} and the scale height $H$ is 
very small, i.e., $H/r$ is typically $\sim$0.01.  
The X-ray flux which the disk surface receives is 
roughly estimated to be less than 
1$\times$10$^{31}$~erg~s$^{-1}$.  
This is smaller than the expected intrinsic disk flux 
in the quiescent state by an order of magnitude 
(see e.g.,~\cite{sma84DI,whe03sscyg}).}
Therefore, the irradiated disk emission would be trivial.  

If the secondary star is irradiated by hard X-rays from 
the BL, its temperature is altered as 
\begin{equation}
{T_{\rm 2, irr}}^4 = {T_2}^4 + (1 - \eta_{\rm S}) L_{\rm X} \frac{1}{4 \pi a^2 \sigma},
\label{T2-irr}
\end{equation}
where $T_{\rm 2, irr}$ and $T_{2}$ are the temperature 
of the irradiated secondary star and 
the intrinsic temperature of the secondary, which is 
estimated to be 4,750~K \citep{hil17sscyg}, respectively 
(see equation (1) of \cite{ham20modeling}).  
Also, $\eta_{\rm S}$ is the albedo of the secondary star, 
which is $\sim$0.4 \citep{dejon96LMXBreprocessing}, 
$L_{\rm X}$ is the X-ray luminosity, 
$a$ is the binary separation, 
and $\sigma$ is the Stefan-Boltzmann constant 
\citep{ham20modeling}.  
\textcolor{black}{By substituting the X-ray luminosity 
estimated from our SED fittings, we see that} 
X-ray irradiation increases the temperature of 
the secondary star by only a few percent at most.  
Consequently, the irradiation of the disk and 
the secondary star would not be the main source of 
the increase in the optical flux.  

As a result of the above considerations, \textcolor{black}{the third 
possibility seems to be most plausible}.  
The accretion rate in the disk would increase, which leads to 
the brightening of the disk.  
According to \citet{ham20modeling}, the optical emission 
from the disk plus the bright spot is about half of 
that from the secondary star during the normal quiescence 
in SS Cyg.  
To explain the optical luminosity during Event A, 
the disk should be brighter by $\sim$1.6 times than 
the secondary star.  
In particular, the inner part of the disk may be brighter 
on the basis of the $B-V$ color during Event A, which is 
investigated in subsection 3.3 
(see also the circles in the left panel of Figure \ref{color}).  
The X-ray luminosity is roughly estimated by using the mass 
accretion rate at the inner disk edge ($\dot{M}_{\rm in}$) 
as follows\footnote{\textcolor{black}{We note that 
this equation may be replaced by other forms if the situation is 
more complex (e.g., the inner disk edge is truncated far from 
the WD surface and/or the truncated disk is covered by 
the X-ray emitting coronal flow.}}: 
\begin{equation}
L_{\rm X} \sim \frac{GM_{1}\dot{M}_{\rm in}}{2 r_{\rm in}}.
\label{xray-lumi}
\end{equation}
If the mass accretion is enhanced at the disk inner edge, 
the X-ray flux rises.  
We can thus understand not only the slow growth of the optical 
flux but also that of the X-ray flux in a consistent manner 
by enhanced mass accretion in the inner disk.

\subsection{Enhanced viscosity in the quiescent disk}

In the previous subsection, we have suggested that 
the increase in the mass accretion rate in the inner 
part of the disk may be responsible for the gradual increase 
in the optical and X-ray flux during Event A.  
We next consider the reason why the accretion rate increases.  
The local accretion rate at a given radius of the disk is 
expressed as $\dot{M}(r) = 2 \pi r \Sigma (-v_r)$.  
Here, $v_r$ is the radial velocity by the viscous diffusion and 
$(-v_r) \sim \alpha_{\rm cool} {c_s}^2 / v_{\rm K}$ in 
the quiescent disk, 
where $\alpha_{\rm cool}$ is the viscous parameter in 
the cool state\footnote{Here, we use the term ``the cool state'' 
instead of ``the quiescent state'' because we discuss about 
the thermal state of the disk.}, $c_s$ is the sound velocity, 
$v_{\rm K}$ is the Keplerian velocity, and $\Sigma$ is 
the surface density.  
\textcolor{black}{
The local accretion rate in the quiescent disk is therefore 
proportional to $\alpha_{\rm cool} r^{3/2} \Sigma_{\rm cool} 
T_{\rm c}$, where $T_{\rm c}$ is the temperature 
at the equatorial plane of the disk.  
Here, $\Sigma_{\rm cool}$ is the surface density in the quiescent disk 
and it is larger than $\Sigma_{\rm min}$ typically by a factor 2.  
Since $\Sigma_{\rm min}$ is approximately proportional to $r$ and 
${\alpha_{\rm hot}}^{-0.8}$ (see e.g., \cite{ham98diskmodel}), 
the mass accretion rate at the inner disk edge in the quiescent 
state is roughly expressed as} 
\begin{equation}
\textcolor{black}{\dot{M}_{\rm in} \propto \alpha_{\rm cool} \alpha_{\rm hot}^{-0.8} T_{\rm c} r_{\rm in}^{2.5}.}  
\label{mdot-in}
\end{equation}
The inner edge of the accretion disk during Event A would be 
comparable with or a little smaller than that in the normal 
quiescence according to our estimates in subsection 3.5.  
Therefore, the viscosity and the temperature would increase 
in the cool state in order to account for the increase 
in the optical and X-ray flux.  
If the viscosity in the cool state is enhanced, the viscous 
heating becomes more efficient, which makes the temperature 
of the disk higher.  
\citet{can93DI} showed that the quiescent level rises with 
the increase in $\alpha_{\rm cool}$ by performing numerical 
simulations.  

The enhancement of viscosity in the cool state affects 
light curves as follows.  
It is considered that $\alpha_{\rm cool}$ should be lower 
than the viscosity in the hot state ($\alpha_{\rm hot}$) 
to reproduce a clear-cut cycle between the quiescent state and 
the outburst state as a result of numerical simulations by 
\citet{sma84DI} and \citet{min85DNDI}.  
\textcolor{black}{
Also, outbursts tend to be triggered at the outer disk 
in the case where $\alpha_{\rm cool}$ increases as 
the radial distance from the WD becomes larger 
\citep{min89quiescenceviscosity}, though outside-in outbursts 
are easily triggered if the mass transfer rate is high even 
without the radial dependence of $\alpha_{\rm cool}$ 
\citep{can88outburst,ich94cycle}.  
If the viscosity in the cool state increases especially 
at the inner disk, 
inside-out outbursts frequently occur and their amplitudes 
become smaller.  
Although it may be difficult to distinguish outside-in and 
inside-out outbursts only by optical light curves \citep{sch03sscyg}, 
some numerical simulations and recent eclipse analyses showed 
that the optical rise of inside-out outbursts tends to be 
slower in comparison with that of outside-in outbursts 
\citep{sma84DI,can86DNburst,cou20exdra}.  
Besides, long outbursts may be suppressed if more disk mass is 
drained to the WD during frequent short outbursts due to efficient 
angular momentum transfer in the disk.}  
In fact, SS Cyg repeated small and slow-rise outbursts 
during Event A and showed no long outbursts 
during $\sim$1.5 yr before 2021 (see the upper panel of 
Figure \ref{multi-overall}).  
This behavior may be the evidence of the temporal enhancement 
of $\alpha_{\rm cool}$.  

It is difficult to understand the reason why the viscosity 
in the cool state is enhanced since it is unclear what 
$\alpha_{\rm cool}$ originates from.  
Magnetohydrodynamic (MHD) turbulence producing the viscosity 
in the hot state is considered to be depressed in the quiescent disk 
due to much greater resistivity \citep{gam98}.  
Here, we consider the reason for the increase in viscosity 
in the case of constant mass transfer rates since 
our observations do not imply that mass transfer 
rates change during Event A.  
\textcolor{black}{
A change in the magnetic field of the secondary star, 
such as magnetic cycles and flaring events, may affect 
the angular momentum transport in the disk 
through turbulence or magnetized outflows and control 
the outburst behavior \citep{sce20magnetic,ola16magneticcycle}.}  
Spiral density waves induced by tidal torques might contribute 
to $\alpha_{\rm cool}$ \citep{ju16MHD}, but  
whether it is applicable to the quiescent disk in SS Cyg is 
in debate.  
Although it is proved that thermal convection can enhance 
the MHD turbulence, this is believed to be limited for 
the hot state \citep{hir14ADconvection}.  
Once the temperature is raised and the resistivity is 
lowered at a local region of the disk, 
the MHD turbulence begins to work and the disk temperature will 
increase because of enhanced dissipation, which further 
lowers the resistivity.  
Considering Fig.~1 of \citet{gam98}, if $\alpha_{\rm cool}$ is 
enhanced and the cool branch of the thermal equilibrium curve 
shifts upwards, the resistivity seems to become weaker and 
the MHD turbulence may be switched on.  
Besides, \citet{san02halleffect} investigated nonideal MHD effects 
and showed that 
magnetorotational instability (MRI) operates if the magnetic 
Reynolds number exceeds the critical value.  
This transition may propagate over a wider region on the viscous 
timescale and the viscosity may be enhanced over 
the entire quiescent disk.  
Also, ionization of the gas at the upper layer of the disk 
may contribute to lower the resistivity, though X-ray 
irradiation is not the main source of the increase in 
the optical flux.  
If the lower resistivity switches on MRI, the mass accretion 
rate is raised and the X-ray flux will increase.  
The irradiation effect then becomes stronger.  
However, it is unknown whether this kind of feedback loop 
and global transitions are feasible.  
This remains one of the future problems.  

\subsection{Nature of the 2021 anomalous event in SS Cyg}

We here consider the origin of Event B in SS Cyg 
on the basis of the discussion in subsections 4.1 and 4.2.  
\textcolor{black}{
As described in subsection~4.1, we do not find 
any positive evidence for the enhancement of mass 
transfer rates during Event A.  
Also, the optical mean luminosity during Event B is 
comparable with that before Event B (see subsection 3.1), 
and hence, the released energy from the disk via accretion 
does not seem to drastically change.  
Small-amplitude outbursts do not frequently appear 
with increasing mass transfer rates according to 
\citet{can93DI}.  
We, therefore, investigate how we can explain the anomalous event 
in 2021 without variations of mass transfer rates.}

\textcolor{black}{
X-ray irradiation of the disk and the secondary star 
is not enough to increase the optical flux by $\sim$2 mag 
when considering the observed X-ray flux 
(see Table \ref{parameter-cevmkl}).}  
The increment of the temperature of the secondary star 
by irradiation would be only $\sim$6\% at most 
in Event B according to equation (\ref{T2-irr}).  
The optical luminosity increases only by $\sim$0.3~mag 
at most by X-ray irradiation.  
Although it is possible that the secondary star is 
irradiated by the accretion disk which is bright at optical 
and near-ultraviolet wavelengths, the disk flux must be 
increased for this situation to occur.  
It is therefore natural to assume that the enhancement of 
viscosity in the cool state, which would be caused 
during Event A, is also related to the mechanism of 
Event B.  

\citet{min85DNDI} pointed out that the cooling and heating 
waves are easily trapped in a local region at which 
the thermal instability is triggered if $\alpha_{\rm cool}$ 
is equal to $\alpha_{\rm hot}$.  In their numerical simulations, 
the inner part of the disk always stays in the outburst state 
and the thermal instability arises at a local region in 
the outer part of the disk.  The resultant light curve shows 
small-amplitude fluctuations.  
This phenomenon seems to be similar to Event B in SS Cyg.  
The $B-V$ color during Event B may imply that the inner part 
of the disk is almost always in the outburst state.  
The $V-I$ color during Event B may suggest that the outer part 
of the disk goes back and forth between the cool state and 
the hot state (see the squares in the left and right panels of 
Figure \ref{color}).  
We suggest that $\alpha_{\rm cool}$ gradually approached 
$\alpha_{\rm hot}$ during the pre-stage in this system, 
causing Event B.  
Event B can be interpreted as consecutive low-amplitude 
outbursts caused by the thermal instability triggered 
at the outer part of the disk.  

The insufficient removal of the mass and angular momentum 
from the disk may be also involved in the cause of 
Event B.  
As discussed in subsection 4.2, the enhancement of 
$\alpha_{\rm cool}$ can increase the quiescent level, 
cause frequent low-amplitude and inside-out outbursts, 
and suppress long outbursts.  
A large amount of the disk mass is normally drained onto the WD 
in the long outburst and the mass accretion is not efficient 
in the short outburst (see Fig.~10 of \cite{kim20tiltdiskmodel}).  
It is considered that a larger amount of mass accumulated 
before the long outburst in 2021.  
The redder $V-I$ color during Event A may suggest there was 
a mass reservoir in the cool outer disk (see the circles 
in the right panel of Figure \ref{color}).  
The disk should have drained a lot of mass onto the WD 
during the long outburst in January 2021.  
However, this long outburst showed a shoulder-like precursor 
at the beginning, which lasted for $\sim$5~d.  
Also, there was no clear plateau stage after the system reached 
the outburst maximum (see subsection 3.1).  
The plateau stage of the long outburst is considered to be 
generated by strong tidal torques exerting the outer disk 
edge when the disk reaches the tidal truncation radius 
and the precursor is created before strong tidal torques 
begin to operate.  
The above-mentioned observational features thus suggest 
that the removal of the angular momentum of the disk 
by tidal torques during this long outburst was inefficient 
and that a lot of mass may remain in the disk even after 
this outburst was quenched around BJD 2459230.  
The remaining mass in the outer disk would be gradually 
consumed and help to maintain Event B.  
\textcolor{black}{
Although the multi-wavelength behavior around BJD 2459100 
was similar to Event B, it was not long-lasting.  
This might be because the amount of the disk mass stored before 
this phenomenon was smaller than that just before Event B.}

\subsection{Standstill-like phenomenon in SS Cyg-type stars and its relation with other types of standstills}

As mentioned in the introduction, Event B at first appeared 
to be the same phenomenon as Z Cam-type standstill; 
however, our analyses suggest that this would be a group 
of small-amplitude outbursts.  
Event B is a standstill-like phenomenon at this stage 
rather than standstill, though the definition may be 
changed after the end of this phenomenon.  
Recently, a number of IW And-type DNe which repeat 
quasi-standstills terminated by small brightening 
have been discovered one after another.  
Here, quasi-standstills are intermediate brightness 
intervals with (damping) oscillations (e.g., 
\cite{kat19iwand} and \cite{lee21hopup}).  
The thermal state of the disk during Event B seems to 
resemble that during quasi-standstills in IW And-type DNe, 
which was investigated by \citet{kim20tiltdiskmodel}, 
rather than that during Z Cam-type standstill.  
According to their work, in IW And-type quasi-standstill, 
the inner disk is always in the hot state and the thermal 
instability is triggered in the outer disk, though it 
is unclear if this picture is true or not in light of 
recent observations \citep{kim20kic940}.  

\begin{figure*}[htb]
\begin{center}
\FigureFile(160mm, 50mm){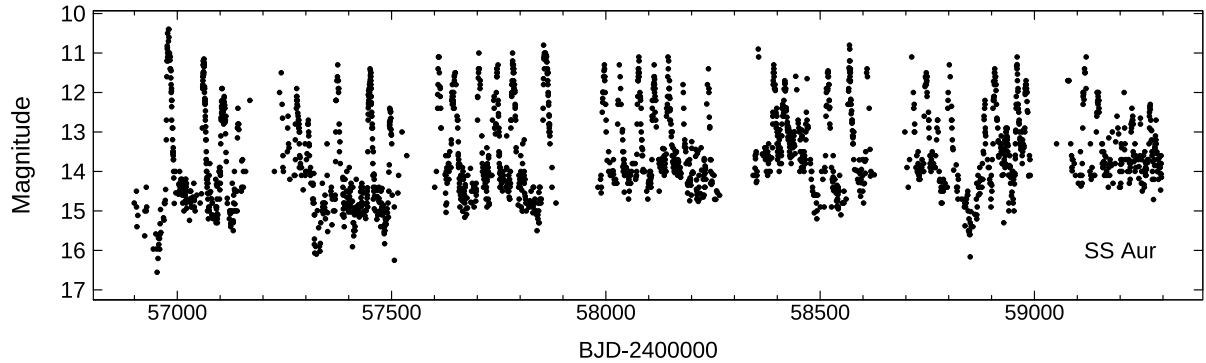}
\end{center}
\caption{
Optical light curves of SS Aur, which are taken from the AAVSO archive.  We here use the photometric data in the clear filter and the $V$ band and the visual observations.  
}
\label{ssaur}
\end{figure*}

We have noticed that some SS Cyg-type stars have entered 
a standstill or may show standstill-like phenomena like SS Cyg.  
For example, WW Cet experienced standstill in 2010.  
Moreover, its forerunner was the gradual increase in the optical 
flux in the quiescent state as in SS Cyg (see the top panel 
of Figure E6).  
Additionally, another bright SS Cyg-type star SS Aur becomes 
more active these days and repeats low-amplitude outbursts.  
We see from Figure \ref{ssaur} that the quiescent flux is 
gradually increasing in this system.  
This system may enter standstill or standstill-like phases 
as SS Cyg and WW Cet did by considering the similarity 
in the precursor events.  
TW Vir would also be in the same state as SS Aur.  
We give in Figure E6 the light curves of some DNe 
which have shown long-lasting increases in the optical flux 
in the quiescent state.  
We suggest that SS Cyg-type stars having mass transfer 
rates lower than the critical rate may show standstill-like 
events if some critical conditions such as an enhancement of 
viscosity in the cool state and a large amount of 
the accumulated mass in the outer disk are satisfied.  

We consider that standstill or standstill-like phenomena 
in CVs are classified into two types: Z Cam-type standstill 
and standstill-like phenomena in SS Cyg-type stars, which 
possibly link to IW And-type quasi-standstills.  
The increase of the quiescent level is seen in the latter, 
and not in the former.  
The entire disk is thermally stable in the former, while 
the outer part of the disk is thermally unstable 
in the latter.  
The critical condition of the former type is high 
mass transfer rates close to $\dot{M}_{\rm crit}$, but 
that of the latter type is not.

\subsection{Evolution of the X-ray emitting region}

We have presented so far our interpretation about 
what happened to the accretion disk during Event A and 
Event B.  
We here try to understand the time evolution of the BL.  
In the case of SS Cyg, the X-ray luminosity drops, 
and the X-ray spectrum becomes softer in the optical outburst 
\citep{whe03sscyg,mcg04sscygXray,ish09sscygSuzaku}.  
This would be because the BL becomes optically thick due to 
a lot of mass provided from the inner disk edge 
\citep{nar93BL,pat85CVXrayemission1,pat85CVXrayemission2}.  
In comparing our results with the observation presented 
by \citet{whe03sscyg}, the hardness ratio was comparable 
with that in the optical quiescence throughout T1--T5.  
This means that the BL was an optically-thin coronal flow 
during Event A and Event B.  
In fact, $\dot{M}_{\rm in}$ estimated by equation (\ref{xray-lumi}) 
is lower than the critical mass accretion rate at the inner 
disk edge ($\dot{M}_{\rm in, crit}$) above which the BL becomes 
optically thick\footnote{According to equation (6.20b) in \citet{war95book}, 
$\dot{M}_{\rm in, crit}$ is around a few times of 
10$^{17}$~g~s$^{-1}$ for SS Cyg.  Here we use the binary 
parameters measured by \citet{hil17sscyg}.}.  
It is known that the X-ray luminosity rises by a factor of 
$\sim$5 just before/after an outburst in SS Cyg probably 
corresponding to $\dot{M}_{\rm in}$ (e.g., \cite{whe03sscyg,mcg04sscygXray}).  
The BL might be barely optically thin in these time periods, 
since the X-ray luminosity became higher by $\sim$5 times in 
T4 and T5 than in the normal quiescence.  

We next attempt to interpret the time evolution of 
the properties of the coronal flow other than its luminosity.  
During the normal optical quiescence, 
the solid angle subtended by the reflector was more than 1 
and the maximum temperature of the plasma was 
around 20~keV \citep{ish09sscygSuzaku,mcg04sscygXray}.  
By contrast, the reflection effect was weak in T1--T5.  
The maximum temperature of the corona was 1.5 times higher 
at T1--T3 and the high-temperature component seems to 
have increased.  
The maximum temperature dropped a little in T4 and T5 and 
the photoelectric absorption increased from T3 to T5.  
In particular, the spectrum at T5 suggests that part of 
the plasma was masked by some dense absorbers 
(see sub-subsection 3.4.2).  
Also, it is suggested that the disk inner edge was 
a little smaller in T1 and T2 in comparison with the estimates 
by \citet{bal12xrayDNe} in the normal quiescence, but 
the inner region of the disk was still truncated.  
The inner disk edge may extend down to the vicinity of the WD 
in T4 and T5 (see subsection 3.5).  

To explain the significant reflection effect during 
the normal quiescence, 
the optically-thin corona is believed to have been compact 
with respect to the WD radius \citep{ish09sscygSuzaku}.  
The weak reflection in T1--T5 suggests that the corona 
greatly expanded especially in the vertical direction during 
Event A and Event B.  
Its scale height should exceed the WD radius and this picture 
is similar to a coronal siphon flow proposed 
by \citet{mey94siphonflow}.  
This would be consistent with an increase in the temperature 
of the corona.  
The higher the temperature of the corona is, the higher 
the gas pressure is, and the larger the scale height is.  
Also, the gas in the inner disk would be ionized during 
Event B, and the reflection effect by neutral material 
would be lower.  

In T1--T3, the maximum temperature and the density 
of higher-temperature gas of the coronal flow increased, 
which can be attributed to two factors.
One is the shrinkage of the inner disk edge.  
If the gas cooling is inefficient, the gas is possibly 
heated up to the virial temperature defined as follows: 
\begin{equation}
T_{\rm vir} \sim \frac{GM_{1} m_{\rm p}}{6 k_{\rm B} r_{\rm in}}, 
\label{virial}
\end{equation}
where $m_{\rm p}$ is the proton mass and $k_{\rm B}$ is 
the Boltzmann constant \citep{hel01book,kat08BHaccretion}.  
The smaller the inner disk edge is, the higher the maximum 
temperature is, \textcolor{black}{
though the situation may not be simple as mentioned in subsection~4.1}.  
The second factor is an enhancement of the viscous 
friction rate in the corona.  
If the viscous heating rate increases, hot components 
in the coronal flow are supposed to increase.  

The maximum temperature of the plasma decreased from T3 to T4.  
This change could be explained if we consider 
the increase in the density of the coronal flow.  
The density would be relatively low 
in T1 and T2 since $\dot{M}_{\rm in}$ was supposed to be 
not very high, and then, the radiative cooling of the corona 
was not efficient.  
The corona can be heated up to the virial temperature 
in this case.  
By contrast, the contraction of the disk inner edge suggests 
that a larger amount of gas was conveyed from the disk 
inner edge to the corona in T4 and T5 a while after 
the system entered Event B.  
More mass might evaporate into the coronal flow from a wider 
area of the inner disk.  
In these time periods, the coronal density would become 
higher and the radiative cooling would become more efficient.  
Under this situation, the maximum temperature is supposed to drop.  
At the beginning of Event B, i.e., at T3, 
the density might not be so high as that at T4 and T5.  

It is difficult to understand what the intrinsic absorption 
of SS Cyg originates from.  
The mass-loss rate due to outflows in non-magnetic CVs is 
about less than 1\% and the outflows are observed only 
in the outburst state \citep{lon02ADwindproc}.  
However, it may give us some implications that the SED modeling 
of the spectra in T5 is improved by using the model for partial 
photoelectric absorption (see also Figure \ref{broad-xray-spectra}).  
As discussed by \citet{don97sscygXray}, part of the X-ray corona 
might be hidden from the observer by the optically-thick disk 
which extends down to the WD surface.  
Although the SEDs in T3 and T4 can be fitted well without 
partial absorption, the absorption by the disk may begin 
developing around T3 and T4.  
The increase in the absorption effect from T3 to T5 may suggest 
that the inner disk became optically thick, maintained 
the outburst state, and extended down to the WD surface 
during Event B.  

\begin{figure}[htb]
\begin{center}
\FigureFile(80mm, 50mm){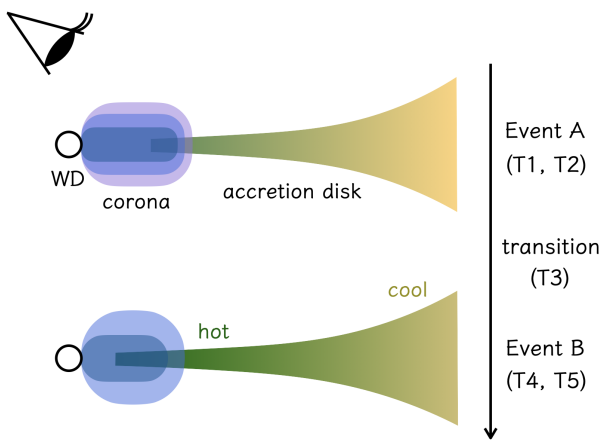}
\end{center}
\caption{
Illustration of the edge-on view of a half of the entire accretion flow during Event A and Event B.  Here, T1, T2, T3, T4, and T5 correspond to the time zones when the coordinated X-ray observations by {\it NICER} and {\it NuSTAR} were performed (see also Figure \ref{broad-xray-spectra}).  The purple region has a higher temperature than the blue region.  
}
\label{accretion}
\end{figure}

We finally illustrate the entire accretion flow of SS Cyg 
during Event A and Event B in Figure \ref{accretion}.  
The corona would be a multi-temperature plasma for all of 
the observations and the corona and the inner part of 
the accretion disk would have a higher temperature than usual.  
The increase in accretion rates of the disk during Event A, 
which was probably caused by enhanced viscosity, might make 
the innermost disk radius contract a little and the maximum 
temperature of the corona became higher.  
Then the coronal flow expanded vertically due to high temperature.  
After T3, the accretion rate to the corona became much higher 
since the inner disk would enter the outburst state.  
As a larger amount of mass was provided to the corona, 
the inner edge of the disk gradually approached the WD surface, 
and the higher-density coronal flow was efficiently cooled down.  
It would be controversial whether the disk is sandwiched by 
the coronal flow.

\section{Summary}

SS Cyg has entered since January 2021 an unusual 
phenomenon possibly similar to standstill observed 
only in Z Cam-type stars.  
We observed this anomalous event and its multi-wavelength 
transition in this source at optical and X-ray wavelengths 
via the VSNET team, the AAVSO, {\it NICER} and {\it NuSTAR}.  
The main results are summarized as follows.  

\begin{itemize}
\item
The optical and X-ray flux gradually and simultaneously 
increased in the optical quiescent state before 
the anomalous event for $\sim$500~d.  
Also, small-amplitude outbursts frequently 
occurred in the pre-stage of the anomalous event.  
\item
We found that the orbital hump was not amplified 
on average during the quiescent state in the pre-stage of 
the anomalous event in 2021.  
\item
The $B-V$ color suggests that the inner part of the disk 
would be hotter than usual during the quiescent state 
in the pre-stage and that the hot inner part might be 
always in the outburst state during the anomalous event.  
The $V-I$ color suggests that a large amount of cool gas 
stored in the outer part of the disk and/or the disk expanded 
during the quiescent state in the pre-stage and during 
the anomalous event.  
\item
The X-ray spectra during the optical quiescent state 
in the pre-stage and during the anomalous event are well 
expressed by a sum of bremsstrahlung emissivity.  
The reflection by the surface of the WD and/or the disk 
was less conspicuous than usual and the temperature of the BL 
became higher.  
The maximum temperature of the BL decreased 
in the middle of the anomalous event.  
The X-ray luminosity were higher by $\sim$2--5 times than usual.  
\item
The innermost disk radius estimated from the PSs of 
X-ray light curves is $\sim$3.5$\times$10$^{9}$~cm during 
quiescence in the pre-stage, which implies that the disk 
was truncated far from the WD surface.  
By contrast, the disk inner edge would extend down to 
the vicinity of the WD during the anomalous event.  
\end{itemize}

\textcolor{black}{
We have investigated the cause of the anomalous phenomenon 
and its predecessor by the disk-instability model 
in the simple case where the mass transfer rate 
does not fluctuate since we have not found any positive evidence 
of a significant increase in mass transfer rates.  
X-ray irradiation was not the main source of the increase 
in the optical quiescent flux during the pre-stage.  
It is suggested that} an increase in mass accretion rates of 
the disk, which would be caused by an enhancement of viscosity 
in the cool state, raised not only optical flux but also 
X-ray flux during the quiescent state.  
We also suggest that the anomalous event in 2021 was 
triggered by the enhancement of viscosity.  
During this event, the inner disk would be always 
in the outburst state and the thermal instability might work 
only in the outer part of the disk.  
The anomalous event in 2021 in SS Cyg would be consecutive 
small outbursts and a standstill-like phenomenon.  
\textcolor{black}{
The combination of a simple case of the disk-instability 
model and increased viscosity in the quiescent disk 
could explain standstill and standstill-like phenomena 
in SS Cyg-type stars.}  
The X-ray emitting inner accretion flow would expand 
spatially because of the increase in its temperature.  
A slight decrease in the maximum temperature of the flow and 
the contraction of the innermost disk radius in the middle 
of the anomalous event \textcolor{black}{seems to be} consistent 
with the increase in accretion rates at the inner disk edge.  
Thus, the change in the accretion rate of the disk 
is likely to influence the structure and the temperature of 
the X-ray emitting corona.

\section*{Acknowledgements}

We are thankful to many amateur observers in the VSNET, 
VSOLJ, and AAVSO and Toshihiko Katayama for providing 
a lot of data used in this research.  
We are grateful to the {\it NuSTAR} team for performing 
the ToO observations.  
We are grateful to Yoji Osaki who gave us a lot of 
insightful comments.  
We appreciate Shigenobu Hirose who commented about 
the origin of viscosity in the quiescent disk.  
M.~Kimura is grateful to Yohei Nishino, Shigeyuki Sako, 
and Daisaku Nogami for discussing the data analyses.  
This work was financially supported by 
Japan Society for the Promotion of Science Grants-in-Aid 
for Scientific Research (KAKENHI) Grant Numbers
JP20K22374 (MK), JP21K03616 (TK), JP20K14521 (KI), 
JP19K14762 (MS), and JP20H01941 (SY).  
M.~Kimura acknowledges support by the Special Postdoctoral 
Researchers Program at RIKEN.  
The work by P.~A.~Dubovsky and I.~Kudzej was supported by 
the Slovak Research and Development Agency under 
the contract No.~APVV-15-0458.  
\textcolor{black}{
We thank the anonymous referee for helpful comments.}

\section*{Supplementary data}

The following supplementary data is available at PASJ online.  
Tables E1--E4 and figures E1--E6.  

\newcommand{\noop}[1]{}



\end{document}